\documentclass[preprint,12pt]{elsarticle}

\usepackage{amssymb}
\usepackage{amsmath}
\usepackage{amsfonts}
\usepackage{graphicx}
\usepackage{multirow}
\usepackage{subcaption}
\usepackage{hyperref}
\usepackage{booktabs}

\usepackage{xcolor}
\newcommand{\edit}[1]{\textcolor{black}{#1}}

\journal{Computers and Fluids}

\begin{document}

\begin{frontmatter}



\title{Effect of Turbulence-Closure Consistency on Airfoil Identification}


\author{Zhen Zhang}
\author{George Em Karniadakis}%
\affiliation{organization={Division of Applied Mathematics, Brown University},
            addressline={170 Hope Street}, 
            city={Providence},
            postcode={02906}, 
            state={RI},
            country={USA}}

\begin{abstract}
We consider an inverse flow problem in which the airfoil shape is identified from its wake signature, namely the velocity field in the wake of a target airfoil. This is an ill-posed problem and highly sensitive to the accuracy and consistency of the employed turbulence closure. We first demonstrate that shape identification based on a single flow condition is ill-posed, whereas incorporating multiple wake signatures obtained at different angles of attack substantially mitigates this ill-posedness. We then compare the inferred geometries obtained using different turbulence closures and find that inconsistencies among the models lead to markedly divergent shapes. 
\edit{Consequently, we directly compare the geometric sensitivities obtained from different models at fixed shapes, and find up to $250\%$ difference among these sensitivities.}
These findings underscore that turbulence-closure consistency is essential for reliable shape identification and further suggest that effective turbulence models must ensure not only accurate predictions but also physically consistent sensitivities—a principle that should guide the development of both classical and data-driven closure models.
\end{abstract}


\begin{highlights}
\item Inverse identification of airfoil shape from wake velocity fields using an adjoint-based RANS framework.
\item Demonstration that using multiple angles of attack alleviates the ill-posedness of inverse shape identification.
\item Systematic comparison of inversely identified airfoils obtained with S--A, $k$--$\omega$ SST, and $k$--$\varepsilon$ turbulence closures.
\item Evidence that turbulence-closure inconsistency leads to order-of-magnitude differences in both shape and functional errors.
\item Introduction of the concept of sensitivity consistency as a complementary criterion to predictive accuracy for turbulence closures.
\end{highlights}

\begin{keyword}
Inverse shape design \sep Airfoil identification \sep Adjoint method \sep Turbulence closure  \sep Model-form uncertainty




\end{keyword}

\end{frontmatter}



\section{Introduction}

Inverse identification of aerodynamic shapes, such as airfoils, from sparse flow measurements downstream of the body represents an interesting problem in fluid mechanics and engineering design. In computational fluid dynamics (CFD), the Reynolds–Averaged Navier–Stokes (RANS) equations combined with turbulence closures remain the most widely used framework for simulating turbulent flows for engineering Reynolds numbers. In this context, we investigate a fundamental and often overlooked issue: the impact of \emph{model discrepancy} in turbulence closures on the accuracy of inverse shape identification.

\textbf{Inverse problem with model discrepancy.}  
Mathematically, in the inverse problem we seek the shape parameters $\theta$ of an object such that the simulated flow field $\mathcal{F}_M(\theta)$ matches the measured data $u_{\text{obs}}$. The operator $\mathcal{F}_M$ denotes the forward model governed by a physical closure model $M$. Ideally, if $\mathcal{F}_M$ coincides with the true physics $\mathcal{F}_{\text{true}}$, the inferred geometry $\theta^*$ satisfies $\mathcal{F}_{\text{true}}(\theta^*) = u_{\text{obs}}$. In practice, however, $\mathcal{F}_M$ is only an approximation—often a RANS model with an imperfect turbulence closure—leading to the so-called \emph{inverse crime of model inconsistency}:
\[
\mathcal{F}_M(\theta^*) = \mathcal{F}_{\text{true}}(\theta_{\text{true}}) + \epsilon
\quad\Rightarrow\quad \theta^* \neq \theta_{\text{true}}.
\]
This discrepancy can bias the inferred geometry even for noiseless data. Classical inverse analyses typically assume a perfect forward model and focus on data noise or regularization strategies (e.g.\ \cite{Tarantola2005InverseProblemTheory, Stuart2010ActaNumerica}), whereas the present study emphasizes the effect of imperfect model physics. In the context of aerodynamic shape identification, we show that the choice of turbulence closure significantly alters the sensitivity $\partial \mathcal{F}_M / \partial \theta$, thereby changing both the optimization trajectory and the recovered geometry.

\textbf{Turbulence closure and model discrepancy.}  
RANS turbulence closures introduce additional constitutive assumptions to approximate the Reynolds stress tensor, typically through eddy-viscosity models such as Spalart–Allmaras (S–A) \cite{SpalartAllmaras1992}, $k$–$\omega$ SST \cite{Menter1994}, or nonlinear stress–strain relations \cite{Craft1996}. These models are calibrated primarily for forward predictive accuracy—matching mean velocity profiles, drag, or lift—rather than for the correctness of sensitivities with respect to geometric or boundary variations. Consequently, two closures that yield nearly identical forward predictions may produce very different gradients, leading to inconsistent inverse or optimization outcomes.  
Recent advances in data-driven closure modeling (e.g.\ \cite{Duraisamy2019, Singh2017}) aim to reduce model-form uncertainty by learning corrections to RANS models, yet few have examined their performance in inverse or adjoint contexts. Since inverse design represents one of the ultimate applications of CFD simulations, turbulence closures should be evaluated not only for predictive fidelity but also for \emph{sensitivity consistency}. This motivates the need for systematic studies that quantify how closure inconsistencies propagate through inverse problems.

\textbf{Inverse shape optimization by the adjoint method.}  
A common approach for aerodynamic shape identification and optimization is to formulate a PDE-constrained optimization problem:
\[
\min_{\theta} J(\mathcal{F}_M(\theta), u_{\text{obs}}),
\]
where $J$ measures the mismatch between simulated and observed flow quantities. Gradient-based methods are particularly attractive due to their scalability with respect to the number of design variables. The adjoint method provides an efficient means to compute the gradient $\mathrm{d}J/\mathrm{d}\theta$ at a cost nearly independent of the number of parameters \cite{Jameson1988, GilesPierce2000}. Adjoint formulations have become standard in aerodynamic optimization, data assimilation, and flow control. However, when the underlying forward model is inconsistent, the computed adjoint gradient reflects the sensitivities of $\mathcal{F}_M$, not of the true physics. This mismatch may result in geometries that perfectly reproduce the data under one closure but deviate substantially under another, revealing the practical consequences of closure inconsistency in inverse design.

\begin{figure}[!h]
    \centering
    \includegraphics[width=0.6\linewidth]{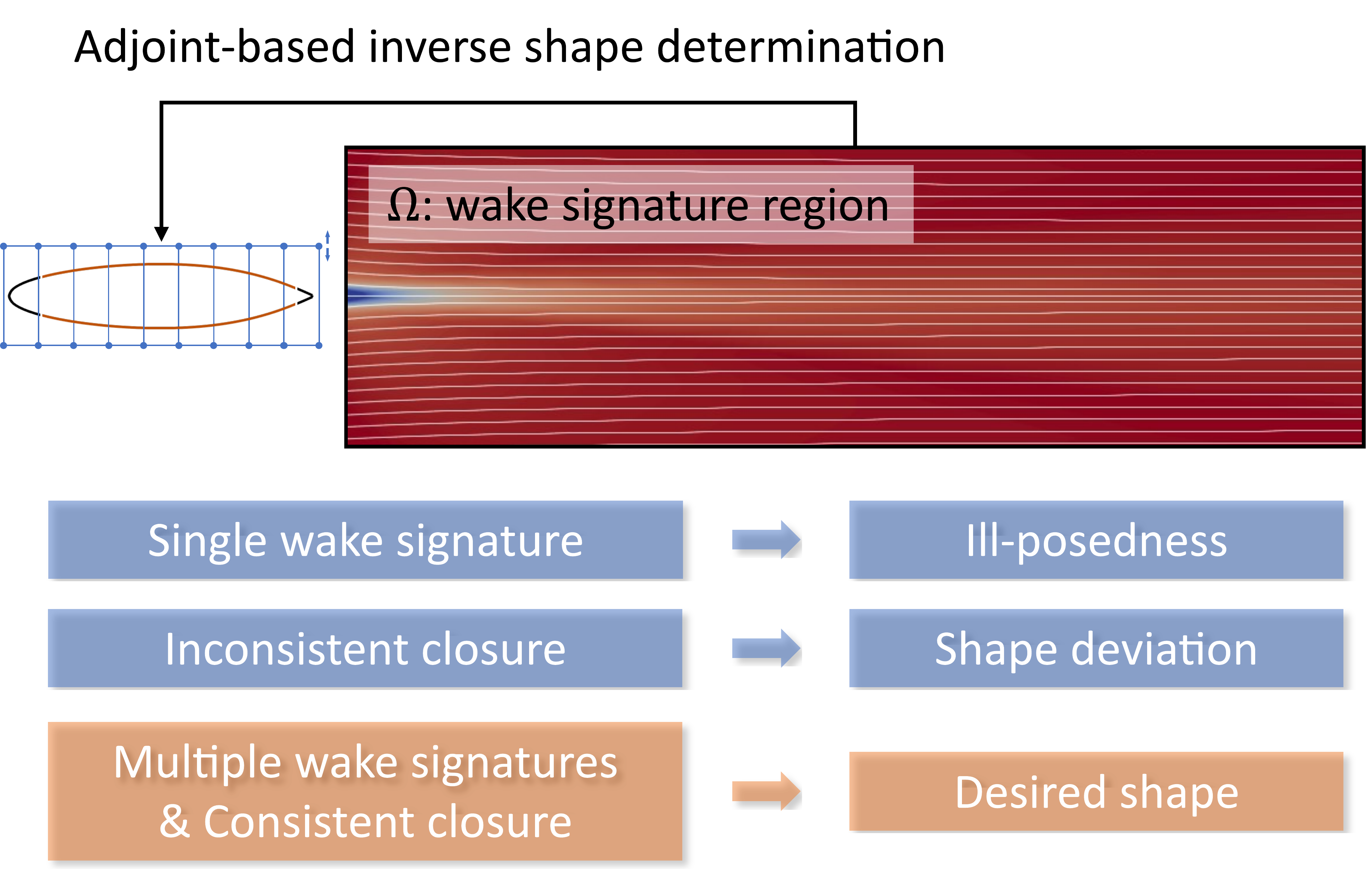}
    \caption{Concept of the present paper. The airfoil shape is identified from the wake signature using the adjoint method. Two issues of the inverse shape identification are discussed: ill-posedness and the effect of the inconsistent turbulence closures.}
    \label{fig:Sketch}
\end{figure}

\textbf{Objectives and contributions of the present study.}  
The present work investigates the inverse shape identification of airfoils from wake velocity fields using RANS equations with various turbulence closures. \autoref{fig:Sketch} summarizes the overall workflow of the study. Specifically, we (i) formulate the inverse problem as a PDE-constrained optimization and demonstrate its ill-posedness under a single flow condition; (ii) show that combining multiple wake signatures at different angles of attack substantially alleviates this ill-posedness; (iii) compare the shapes inferred using several turbulence closures to quantify the influence of closure inconsistency on inverse reconstruction, \edit{and (iv) direct compare the geometric sensitivities obtained from the different closure models at fixed shapes.} Our results highlight that turbulence models must be evaluated not only by their predictive accuracy but also by the \emph{consistency of their sensitivities}. We further discuss two aspects of model discrepancy—functional accuracy and sensitivity consistency—and emphasize their equal significance in inverse design. This finding suggests that future developments of both classical and data-driven closures should target accuracy in both aspects and include uncertainty quantification for each.

\section{Problem Setup and Method}
In this section, we formulate the inverse shape-identification problem and present the corresponding adjoint-based shape-optimization framework, together with the geometric constraints employed.
\subsection{Shape identification from wake signature}
We formulate the following inverse shape identification problem. 

\edit{
\textbf{Objective function.}
The inverse design objective is constructed from three components: the discrepancy in the wake velocity field, the drag coefficient, and the lift coefficient. For a single operating condition, the objective is defined as
\begin{equation}
J
=
w_0\|\mathbf{u}-\mathbf{u}_{obs}\|_{L_2(\Omega)}^2
+
w_1 (C_D-C_{D,obs})^2
+
w_2 (C_L-C_{L,obs})^2,
\label{eq:obj}
\end{equation}
where $\Omega$ denotes the wake region of the airfoil. The weights are selected as
\[
w_0=5\times 10^3, \qquad w_1=1\times 10^7, \qquad w_2=1\times 10^4,
\]
so that the wake-velocity, drag, and lift mismatch terms contribute at comparable magnitudes. In particular, for the baseline NACA0012 airfoil at each of the three Angles of Attack (AoAs) ($0^\circ,5^\circ,10^\circ$), all three terms in \autoref{eq:obj}, excluding the $C_L$ difference at $AoA=0^\circ$, are approximately of order $\mathcal{O}(10)$.
}

\edit{
For the multi-condition case, the objective is extended by summing the mismatch over all operating conditions:
\begin{equation}
J
=
\sum_{m=1}^{3}
\left[
w_0\|\mathbf{u}^{(m)}-\mathbf{u}_{obs}^{(m)}\|_{L_2(\Omega)}^2
+
w_1 \left(C_D^{(m)}-C_{D,obs}^{(m)}\right)^2
+
w_2 \left(C_L^{(m)}-C_{L,obs}^{(m)}\right)^2
\right].
\label{eq:obj_reply}
\end{equation}
This formulation enforces consistency of the recovered shape across multiple flow conditions.
}

\textbf{Shape parameterization and optimization.}
The airfoil shape is initialized from the classical NACA0012 profile and parameterized using Free-Form Deformation (FFD). The design variables are the FFD control points, which are iteratively updated through an adjoint-based optimization procedure to minimize the objective function defined in \autoref{eq:obj} for the single-condition case, or \autoref{eq:obj_reply} for the multi-condition case.

\subsection{Forward simulation}
\begin{figure}[!h]
    \centering
    \includegraphics[width=0.7\linewidth]{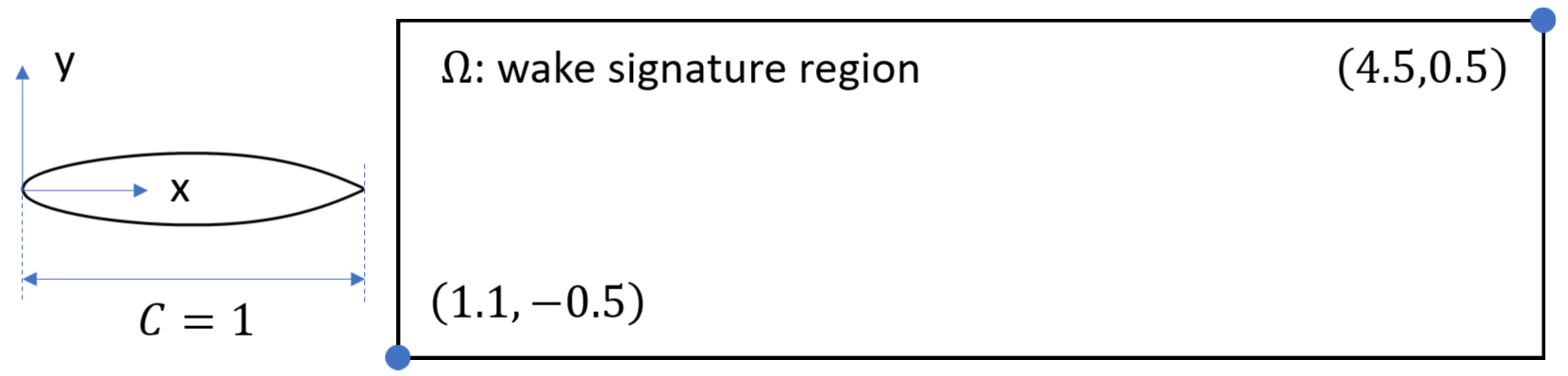}
    \caption{Problem sketch. The origin point is located at the airfoil leading edge. The airfoil chord length is 1. The domain $\Omega \,(x\in[1.1,4.5], y\in[-0.5,0.5])$ is where the wake signature is taken from and the $L_2$ error in \autoref{eq:obj} is calculated.}
    \label{fig:domain}
\end{figure}

\autoref{fig:domain} shows the position of the airfoil and the domain where the wake signature is taken from. The forward simulations of the airfoil flow were performed using the open-source finite-volume solver \textsc{OpenFOAM}. The governing equations are the incompressible, steady Reynolds-averaged Navier--Stokes (RANS) equations closed with selected turbulence models. Pressure--velocity coupling was handled by the SIMPLEC algorithm, and all equations were discretized using a second-order upwind scheme for the convective terms and a central-difference scheme for the diffusive terms. The flow Reynolds number based on the chord length was set to $Re=1\times10^{5}$. Convergence was monitored through the normalized residuals and the stabilization of aerodynamic coefficients, with a typical tolerance of $10^{-7}$ for all variables.

\subsection{Adjoint shape optimization}
The discrete adjoint method is used to solve the PDE-constrained optimization problem.
\begin{equation}
\begin{aligned}
    &\min_\theta \quad & &J(\mathbf{u}) \\
    &\text{subject to} \quad & & \mathbf{R}(\mathbf{u,x(\boldsymbol{\theta})})=0,
\end{aligned}
\label{eq:problem}
\end{equation}
where $J$ is the objective defined in \autoref{eq:obj}, $\mathbf{u}$ is state variable, $\mathbf{x}$ is the vector of mesh coordinates, $\mathbf{R}$ is the residual of the discrete governing equation, and $\boldsymbol{\theta}$ is the design variable in FFD. The PDE constrain in \autoref{eq:problem} makes $J$ a hidden function of the design variable $\boldsymbol{\theta}$, and we can use the Lagrange multiplier method to calculate the total derivative $\frac{dJ}{d\boldsymbol{\theta}}$. The Lagrangian  of this problem is
\[
\mathcal{L}=J(\mathbf{u}) + \boldsymbol{\lambda}^T\cdot \mathbf{R}.
\]
Differentiating Lagrange $\mathcal{L}$ with respect to state variable $\mathbf{u}$ leads to the adjoint equation
\begin{equation}
\left(\frac{\partial \mathbf{R}}{\partial \mathbf{u}}\right)^T \boldsymbol{\lambda} = -\frac{\partial J}{\partial \mathbf{u}}.
\label{eq:adj}
\end{equation}
After solving the adjoint equation and obtaining $\boldsymbol{\lambda}$, we can calculate the total derivatives
\[
\frac{dJ}{d\boldsymbol{\theta}} = \boldsymbol{\lambda}^T \frac{\partial \mathbf{R}}{\partial \mathbf{x}}\frac{d\mathbf{x}}{d \boldsymbol{\theta}}.
\]
Finally, we can update the FFD design variable $\boldsymbol{\theta}$ by passing $\boldsymbol{\theta}\mapsto J,\frac{dJ}{d\boldsymbol{\theta}}$ to the selected optimizers. In this work, the internal point optimizer (IPOPT) is used while the geometric constrains shown in \autoref{sec:constrain} are applied.

Here we adopt the differentiable numerical solver DAFaom~\cite{DAFoamPaper1,DAFoamPaper2} to solve the adjoint equation (\autoref{eq:adj}) by a Jacobian-free Krylov method~\cite{adjoint} as well as automatic differentiation. The derivative $\frac{d\mathbf{x}}{d \boldsymbol{\theta}}$ is obtained by FFD and OpenFOAM's mesh motion solver. 

\subsection{Geometry parametrization and constrains}\label{sec:constrain}
\begin{figure}[!h]
    \centering
    \includegraphics[width=0.5\linewidth]{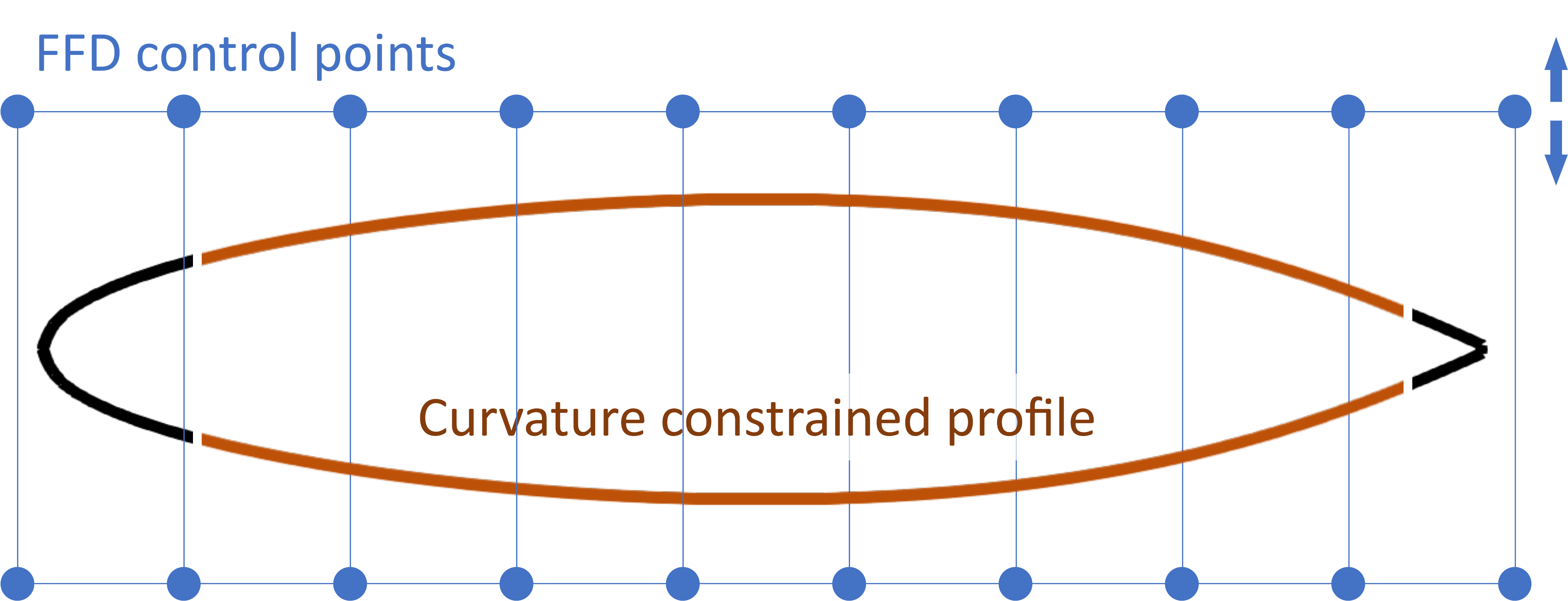}
    \caption{FFD control points and geometry constraints on the airfoil. The FFD control points are designed to move in the y-direction only and keep symmetry along $y=0$.}
    \label{fig:FFD}
\end{figure}

The airfoil geometry was parameterized using the free-form deformation (FFD) technique, in which the baseline airfoil is embedded within a structured lattice of ($2\times 10$) control points, as shown in~\autoref{fig:FFD}. The motion of these control points, defined as the design variables, provides a smooth and flexible deformation field while maintaining geometric continuity and mesh consistency. This parameterization offers an efficient and compact design space that is well-suited for adjoint-based optimization.

To ensure that the optimized geometries remain physically meaningful and manufacturable, several geometric constraints were imposed. \autoref{tab:constraints} summarizes the allowable range and target values of these constraints.
\begin{itemize}
  \item \textbf{Symmetry:} We enforce geometric symmetry about the $x$-axis by displacing each pair of FFD control points at the same $x$-position in opposite directions with equal magnitude.
  \item \textbf{Relative volume constraint ($C_v$):} We maintain the overall volume (area in this 2D problem) of the airfoil relative to the baseline, preventing extreme global shrinkage or expansion during optimization.
  \item \textbf{Pointwise relative thickness constraint ($\mathbf{C_t}\in\mathbb{R}^{10}$):} We calculated it on 10 uniformly distributed point along the chord direction, enforcing the local airfoil thickness relative to the baseline to remain within a prescribed range, ensuring realistic aerodynamic profiles and avoiding excessive camber.
  \item \textbf{Relative leading-edge radius constraint ($C_r$):} We preserve the leading-edge curvature to avoid unphysical sharp or overly blunt leading edges that could significantly alter the boundary-layer characteristics.
  \item \textbf{Mean-square surface curvature constraint ($C_c$):} We calculated it on 20 uniformly distributed points on part of the profile shown in~\autoref{fig:FFD}, limiting high-frequency surface undulations and maintaining geometric smoothness, thereby improving aerodynamic quality and ensuring mesh quality during deformation.
\end{itemize}

\begin{table*}[h!]
\centering
\caption{Summary of geometric constraints and their target values. 
All constraints are defined relative to the baseline NACA0012 airfoil, and the corresponding values for the target NACA16021 airfoil are shown. 
The thickness constraint $\mathbf{C_t}$ is a vector quantity, and the specified range applies to each component.}
\label{tab:constraints}
\setlength{\tabcolsep}{14pt} 
\begin{tabular}{ccc}
\hline
\textbf{Constraint} & \textbf{Allowable range} & \textbf{Target (NACA~16021)} \\
\hline
$C_v$           & $1.5\,{\text{--}}\,3.0$                 & $1.9$ \\
$\mathbf{C_t}$  & $1.2\,{\text{--}}\,4.0$                 & $1.3\,{\text{--}}\,3.0$ \\
$C_r$           & $1.0\,{\text{--}}\,2.0$                 & $1.7$ \\
$C_c$           & $0\,{\text{--}}\,10.0$                  & $3.7$ \\
\hline
\end{tabular}
\end{table*}

\section{Results}
We investigate the inverse shape identification problem under several different setups. \edit{First, in \autoref{sec:mesh_convergence}, we perform a mesh convergence study and determine an appropriate mesh resolution for the subsequent analyses.} Next, in \autoref{sec:multicase}, we compare the inverse problem under single- and multi-condition settings, showing that incorporating additional flow conditions helps alleviate the ill-posedness of the inverse shape identification problem. 
\edit{Subsequently, in \autoref{sec:effect}, we examine the effects of the wake-signature domain size, the objective-function weights, and the number of FFD control points on the identified shape.} 
Then, in \autoref{sec:closure}, we compare inverse solutions obtained with turbulence closures that are either consistent or inconsistent with the model used to generate the wake signature, demonstrating that closure-model consistency has a significant impact on the recovered airfoil shape. 
\edit{Finally, in \autoref{sec:sens}, we compare the flow mismatch and geometric sensitivities predicted by different closure models for fixed shapes, and show that sensitivity inconsistency is a key source of the deviation in the identified shapes.}

\subsection{\edit{Mesh Convergence Verification}}
\label{sec:mesh_convergence}
\edit{
A mesh convergence study is performed for the NACA\,16-021 airfoil at three angles
of attack ($\alpha = 0^\circ$, $5^\circ$, and $10^\circ$) to verify that the
computational mesh provides grid-independent results.  Four systematically refined
multi-block structured meshes are used: \emph{coarse} (7\,220 cells), \emph{medium} (28\,500 cells), \emph{fine} (63\,840 cells), and \emph{finest} (114\,000 cells).
}

\edit{
Drag coefficient $C_D$ and the wake momentum thickness are used as the performance matrices, where the wake momentum thickness is defined as
\begin{equation}
  \theta(x) = \int_{-\infty}^{\infty} \frac{U(x,y)}{U_\infty}
              \left(1 - \frac{U(x,y)}{U_\infty}\right) \mathrm{d}y,
\end{equation}
and is evaluated at four downstream stations $x/c = 1.5$, $3.0$, $5.0$, and
$9.0$.
}

\edit{
The observed order of convergence $p$ is estimated from three successive grids
using
\begin{equation}
  p = \frac{\ln\!\bigl|(f_3 - f_2)/(f_2 - f_1)\bigr|}{\ln r},
\end{equation}
where $f_1$, $f_2$, $f_3$ are solutions on the finest, medium, and coarse grids,
respectively, and the refinement ratio is $r=2$ in each direction.
}

\begin{figure}[h!]
    \centering
    \begin{subfigure}[t]{0.45\textwidth}
        \centering
        \includegraphics[width=\linewidth]{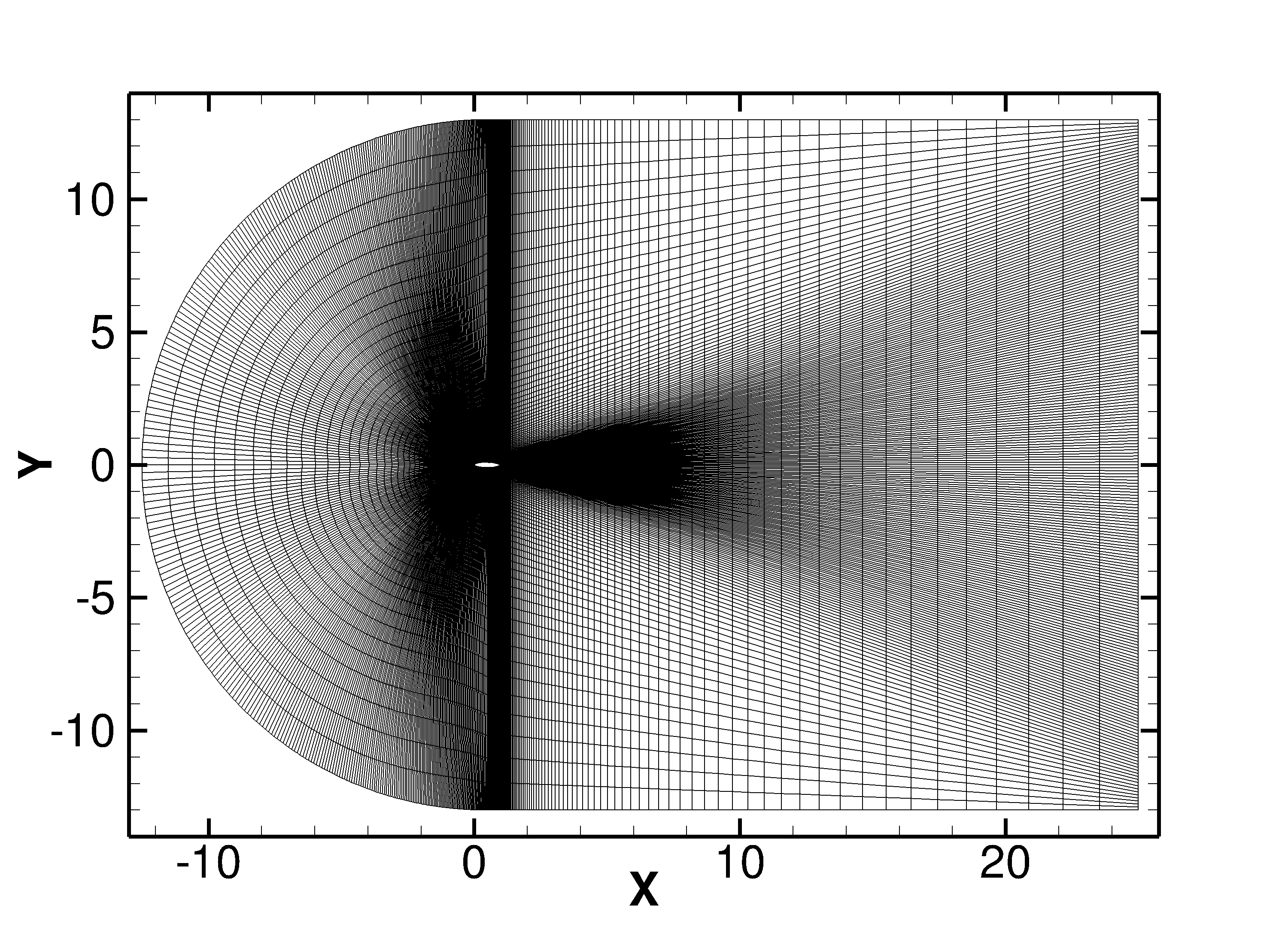}
        \caption{Full view}
    \end{subfigure}
    \begin{subfigure}[t]{0.45\textwidth}
        \centering
        \includegraphics[width=\linewidth]{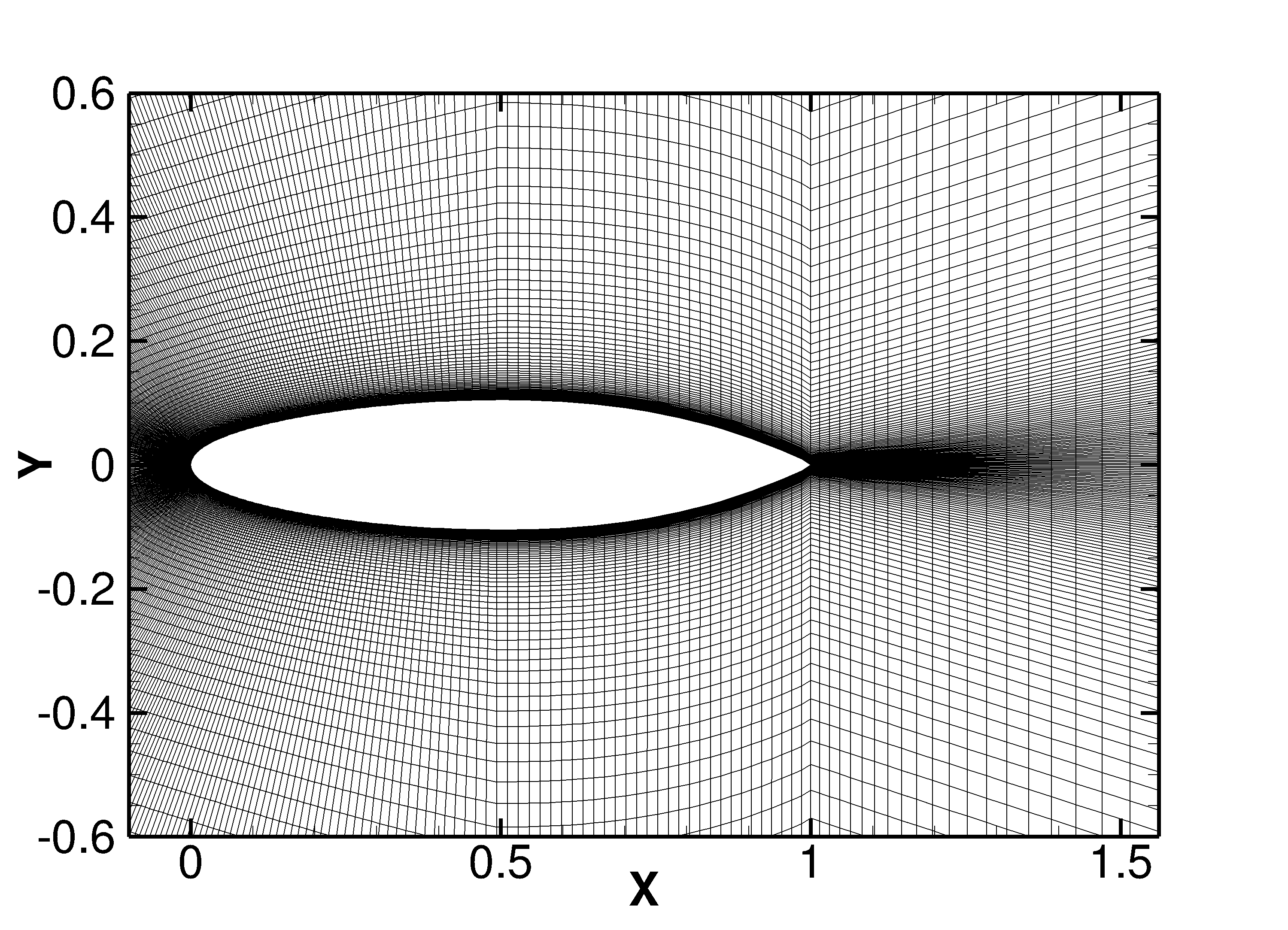}
        \caption{Zoom in view}
    \end{subfigure}
    \caption{\edit{Mesh for NACA16021, fine level, $63\,840$ cells in total.}}
    \label{fig:mesh}
\end{figure}

\edit{
The observed convergence orders for $C_D$ of three angles of attack are $p_{0}=1.87, p_5=2.62, p_{10}=2.83$, and all relative errors against the finest mesh are listed in Table~\ref{tab:mesh_conv_error}. The mesh of the fine level, whose errors are less than $2\%$, is used in the following analysis and is shown in \autoref{fig:mesh}.  
}

\begin{table}[!h]
\centering
\caption{\edit{Relative error (\%) of $C_D$ and wake momentum thickness at each mesh level. Reference values are the finest mesh results.}}
\label{tab:mesh_conv_error}
\begin{tabular}{l l c c c c c}
\hline
$\alpha$ & Level & $C_D$ & $\theta_{x/c=1.5}$ & $\theta_{x/c=3.0}$ & $\theta_{x/c=5.0}$ & $\theta_{x/c=9.0}$ \\
\hline
\multirow{4}{*}{$0^\circ$} & coarse & 29.7 & 27.9 & 16.1 & 15.9 & 19.4 \\
 & medium & 6.4 & 6.0 & 2.9 & 3.5 & 4.4 \\
 & \textbf{fine} & \textbf{1.7} & \textbf{1.5} & \textbf{1.0} & \textbf{0.8} & \textbf{1.4} \\
\hline
\multirow{4}{*}{$5^\circ$} & coarse & 25.0 & 5.0 & 4.8 & 3.7 & 2.8 \\
 & medium & 3.5 & 1.8 & 1.8 & 1.5 & 0.9 \\
 & \textbf{fine} & \textbf{0.7} & \textbf{0.7} & \textbf{0.6} & \textbf{0.5} & \textbf{0.3} \\
\hline
\multirow{4}{*}{$10^\circ$} & coarse & 21.8 & 7.7 & 7.3 & 6.4 & 5.4 \\
 & medium & 2.8 & 2.7 & 2.7 & 2.4 & 2.0 \\
 & \textbf{fine} & \textbf{0.4} & \textbf{1.0} & \textbf{0.9} & \textbf{0.8} & \textbf{0.7} \\
\hline
\end{tabular}
\end{table}

\subsection{Identifying shapes using single and multiple flow conditions}\label{sec:multicase}
The inverse shape identification is an ill-posed problem, which means there may exist multiple solutions to the optimization problem. Here we show that using more information, more concretely, wake signatures from multiple angles of attack can alleviate the ill-posedness of the inverse problem. It is a simple but effective way. 

\begin{figure*}[!h]
\centering
\begin{subfigure}[t]{0.32\textwidth}
    \centering
    \includegraphics[width=\linewidth]{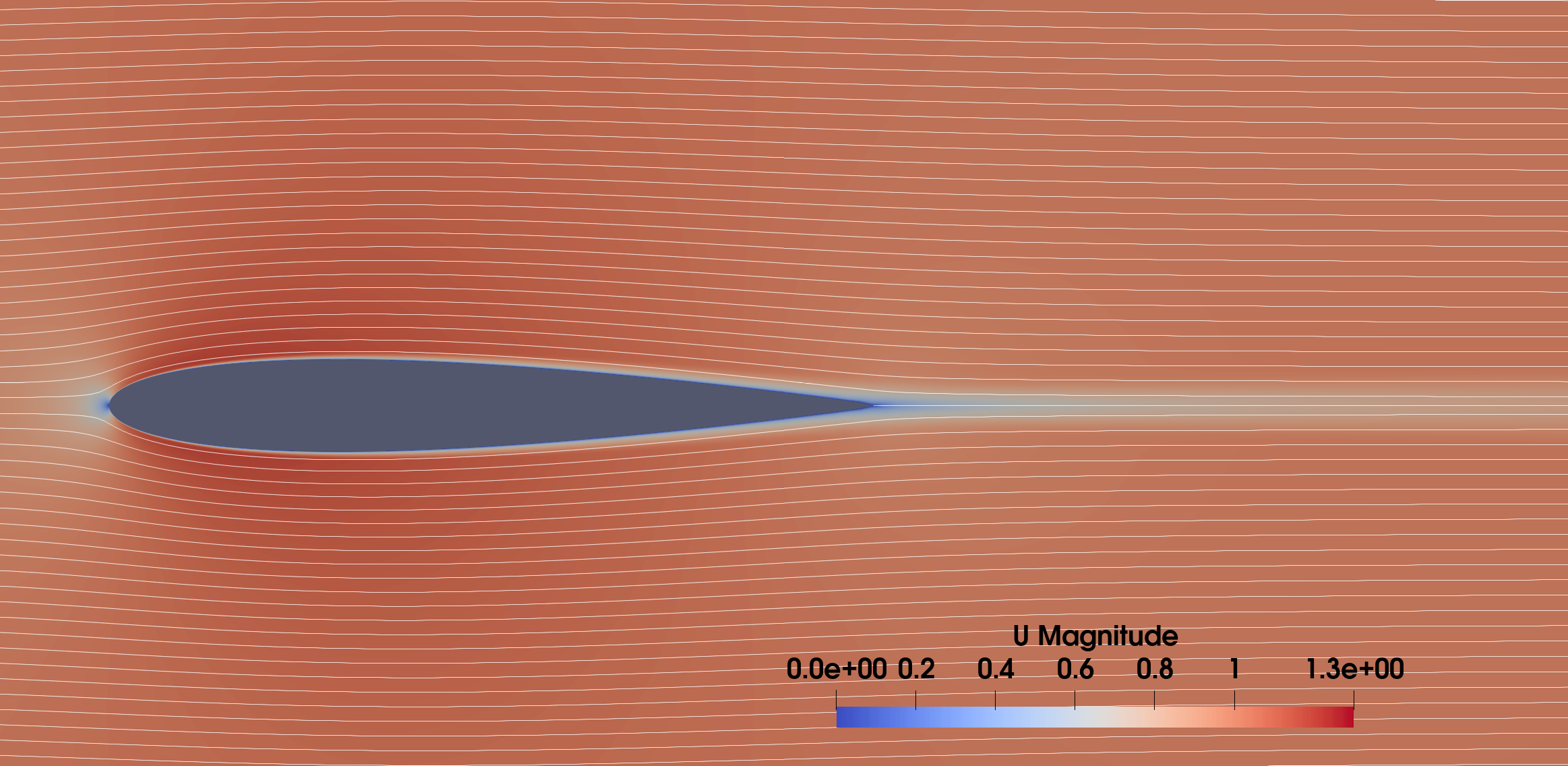}
    \caption{ NACA0012, AoA = 0°}
\end{subfigure}
\begin{subfigure}[t]{0.32\textwidth}
    \centering
    \includegraphics[width=\linewidth]{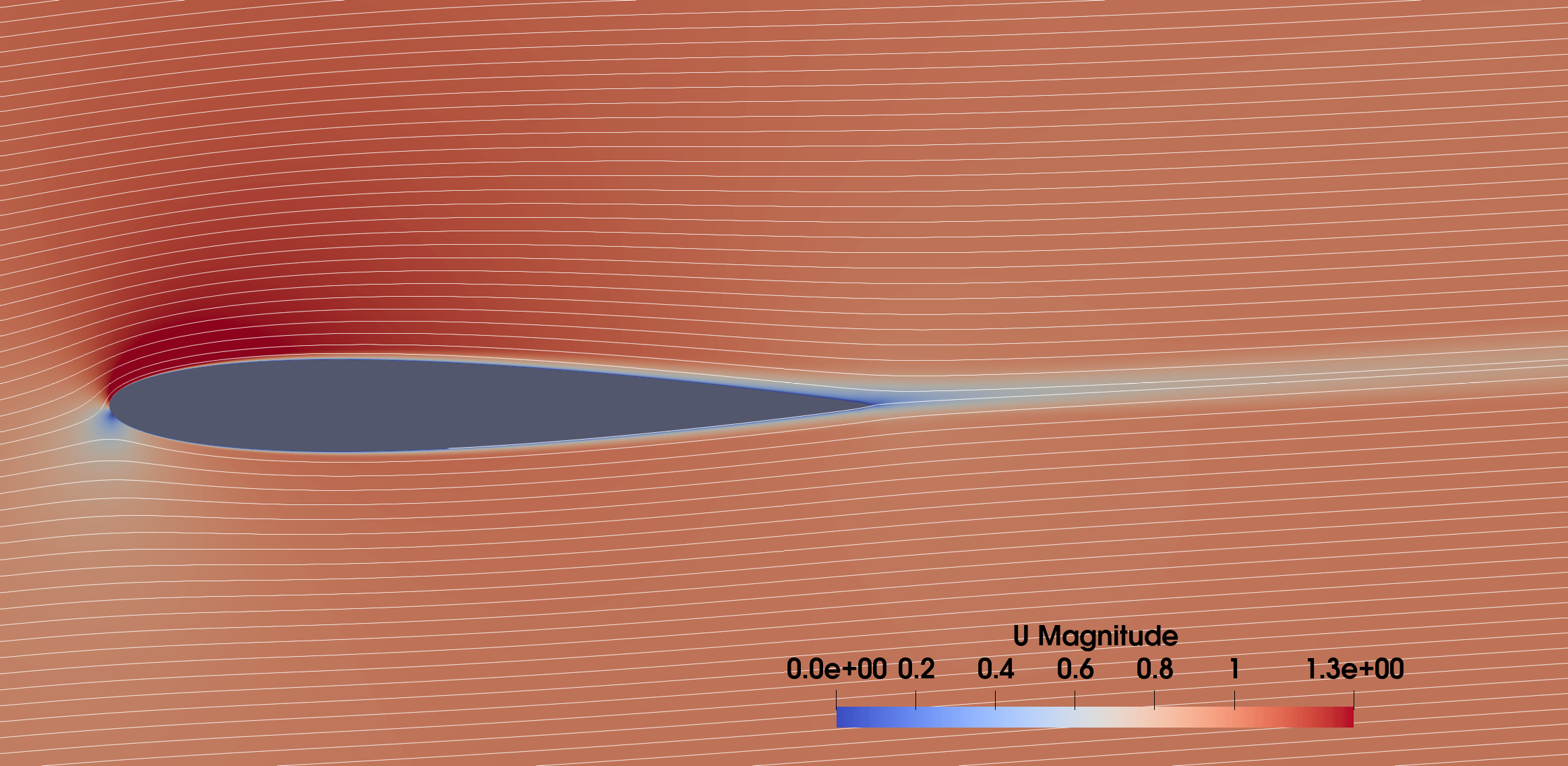}
    \caption{ NACA0012, AoA = 5°}
\end{subfigure}
\begin{subfigure}[t]{0.32\textwidth}
    \centering
    \includegraphics[width=\linewidth]{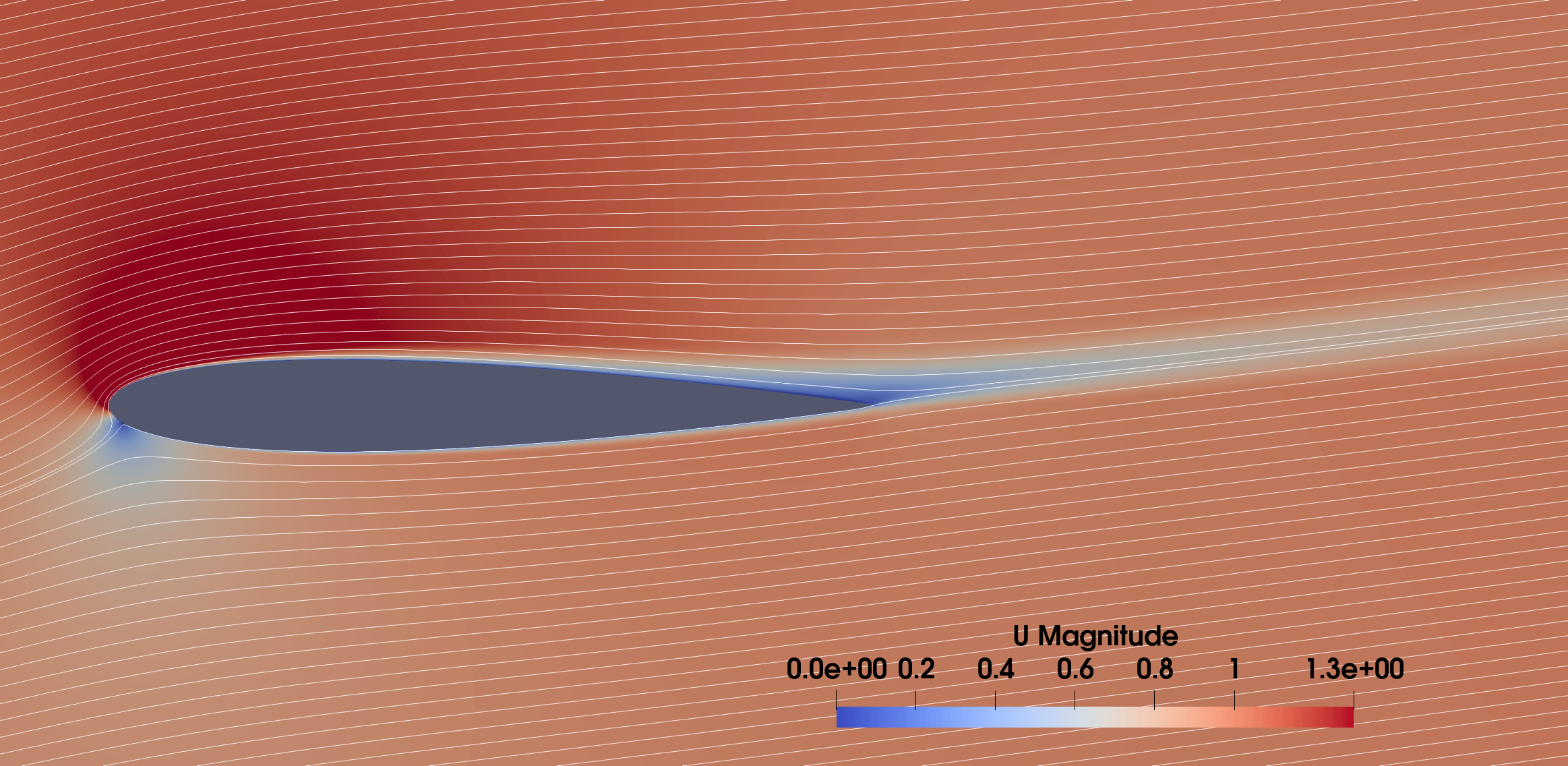}
    \caption{ NACA0012, AoA = 10°}
\end{subfigure}
\begin{subfigure}[t]{0.32\textwidth}
    \centering
    \includegraphics[width=\linewidth]{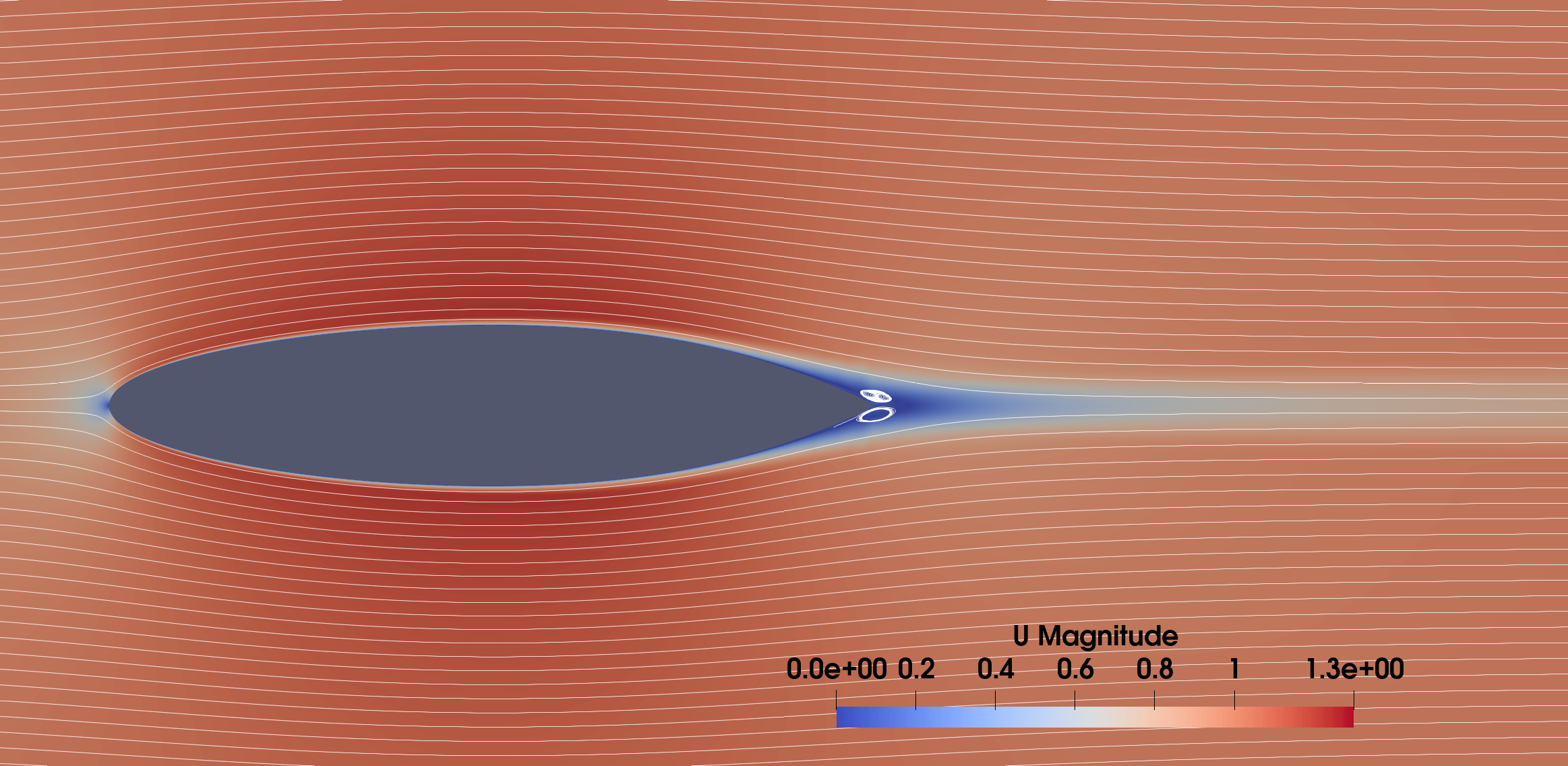}
    \caption{ NACA16021, AoA = 0°}
\end{subfigure}
\begin{subfigure}[t]{0.32\textwidth}
    \centering
    \includegraphics[width=\linewidth]{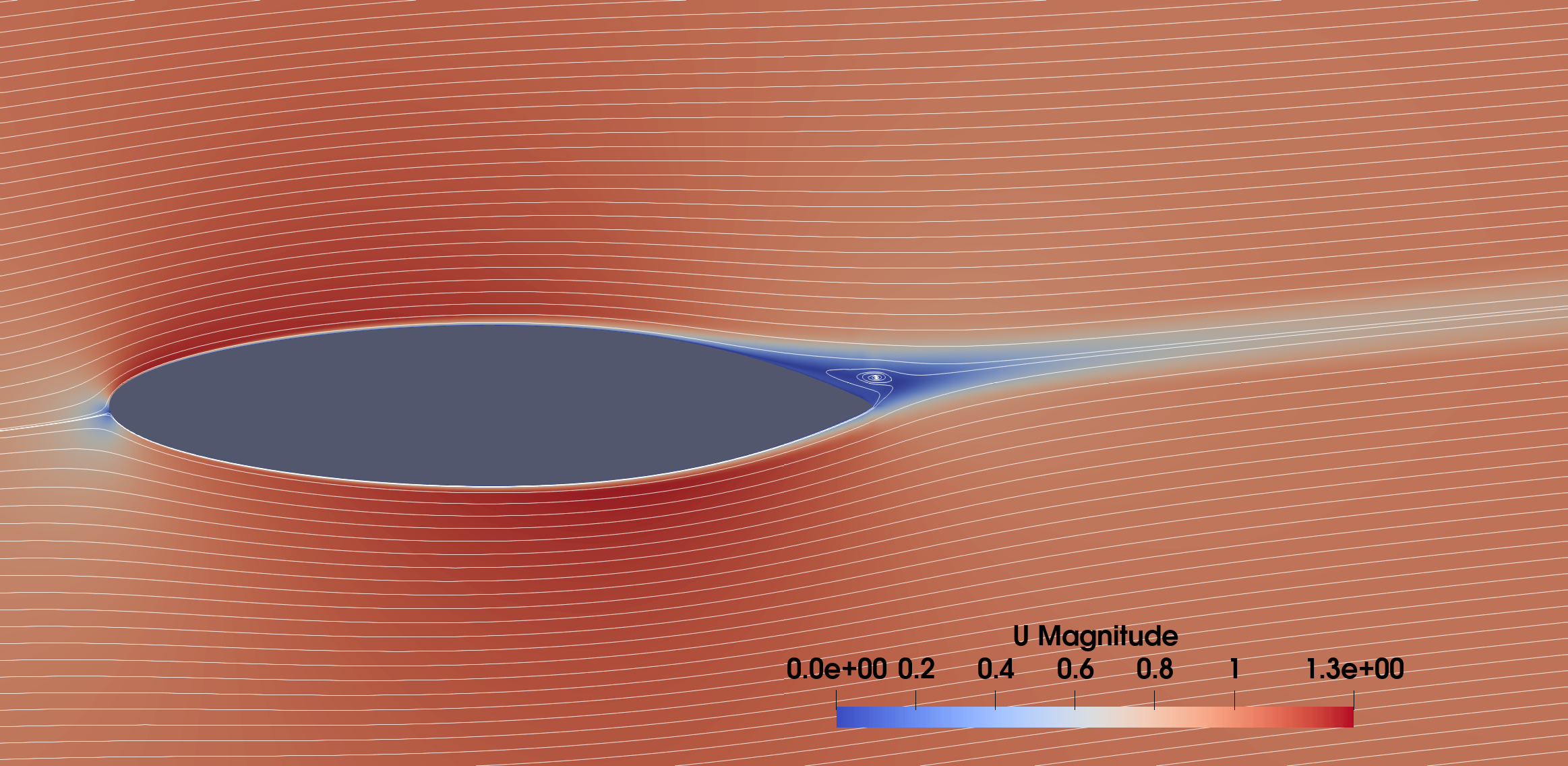}
    \caption{ NACA16021, AoA = 5°}
\end{subfigure}
\begin{subfigure}[t]{0.32\textwidth}
    \centering
    \includegraphics[width=\linewidth]{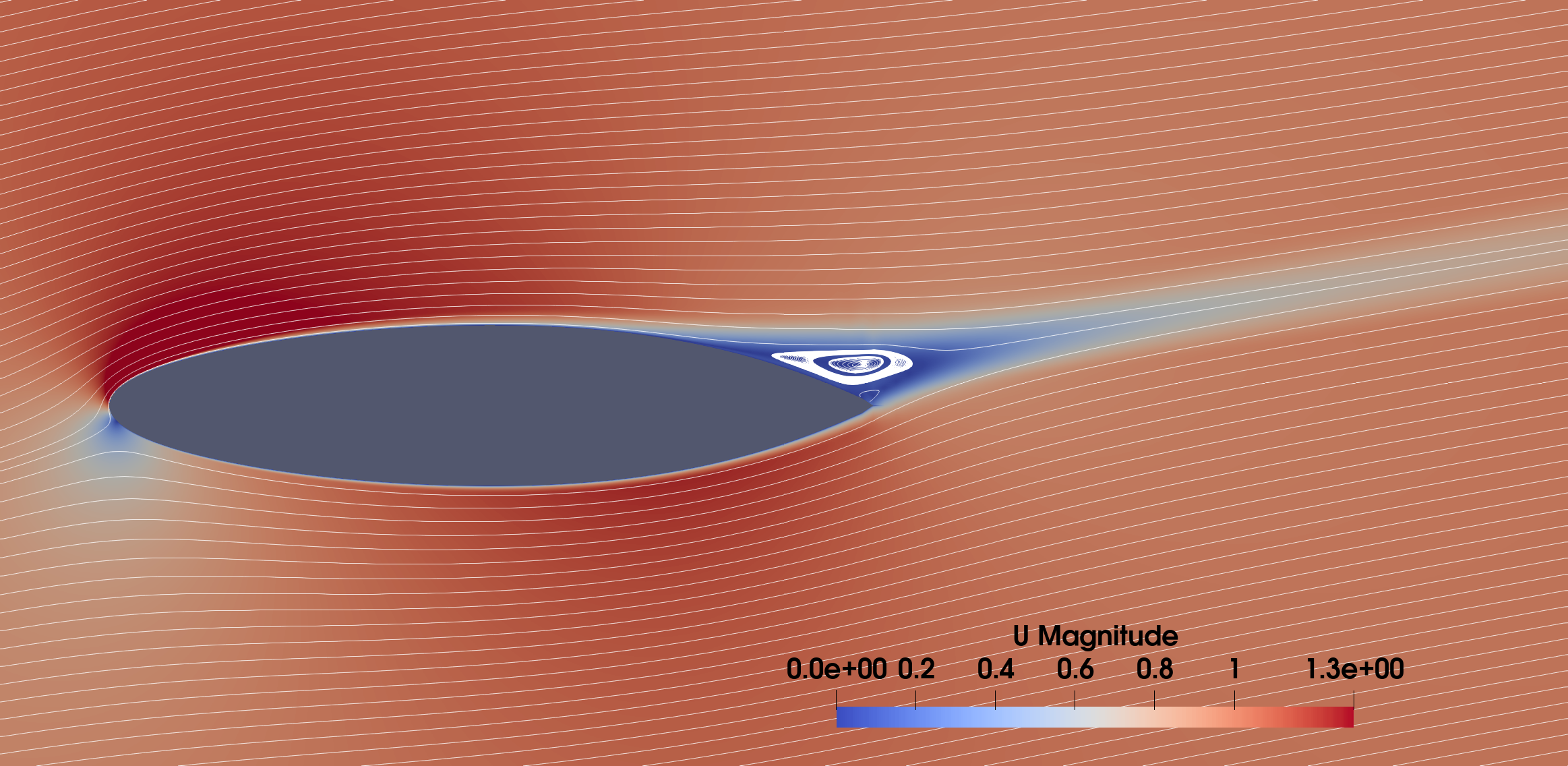}
    \caption{ NACA16021, AoA = 10°}
\end{subfigure}
\begin{subfigure}[t]{0.32\textwidth}
    \centering
    \includegraphics[width=\linewidth]{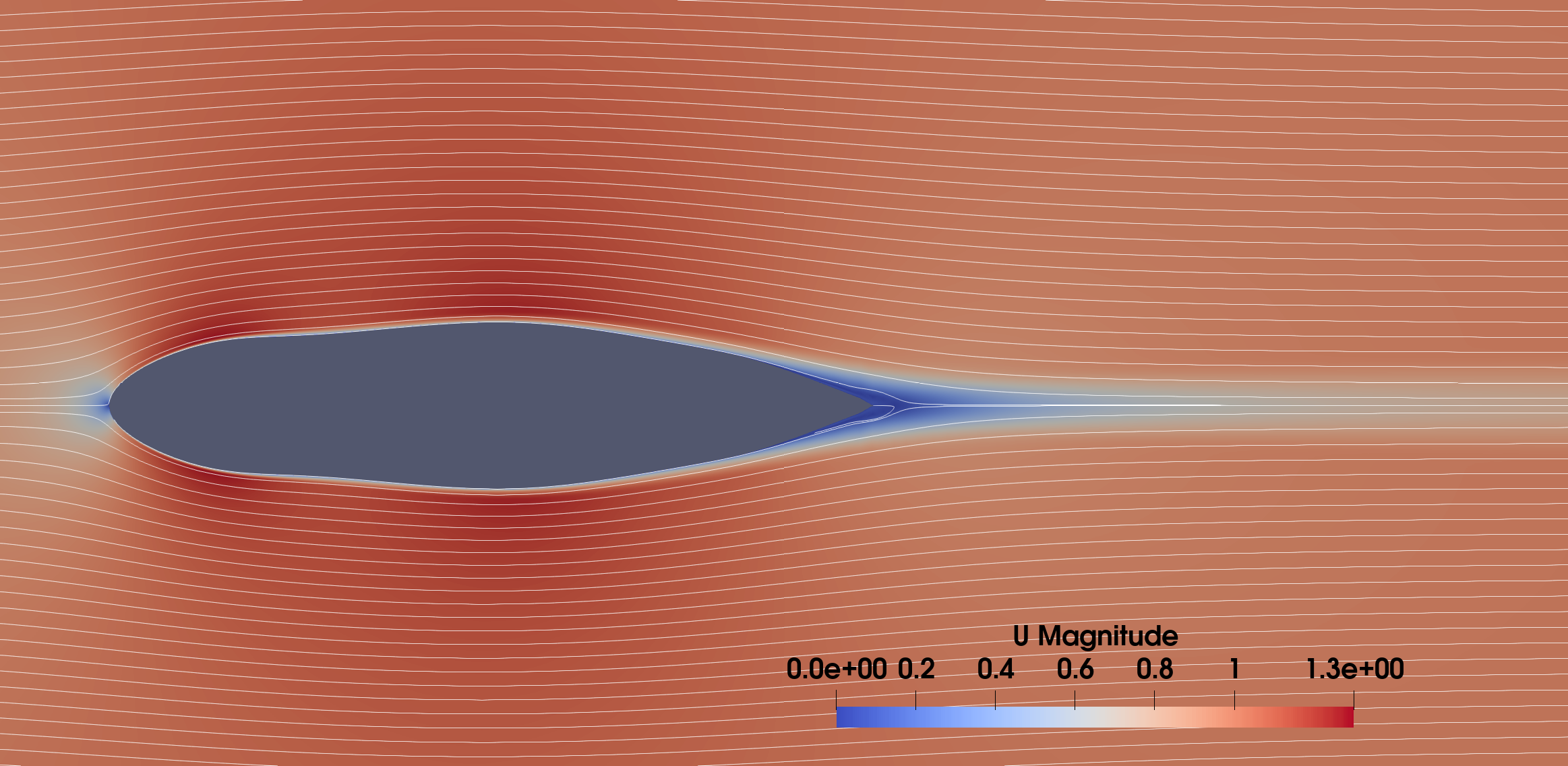}
    \caption{ Single, AoA = 0°}
\end{subfigure}
\begin{subfigure}[t]{0.32\textwidth}
    \centering
    \includegraphics[width=\linewidth]{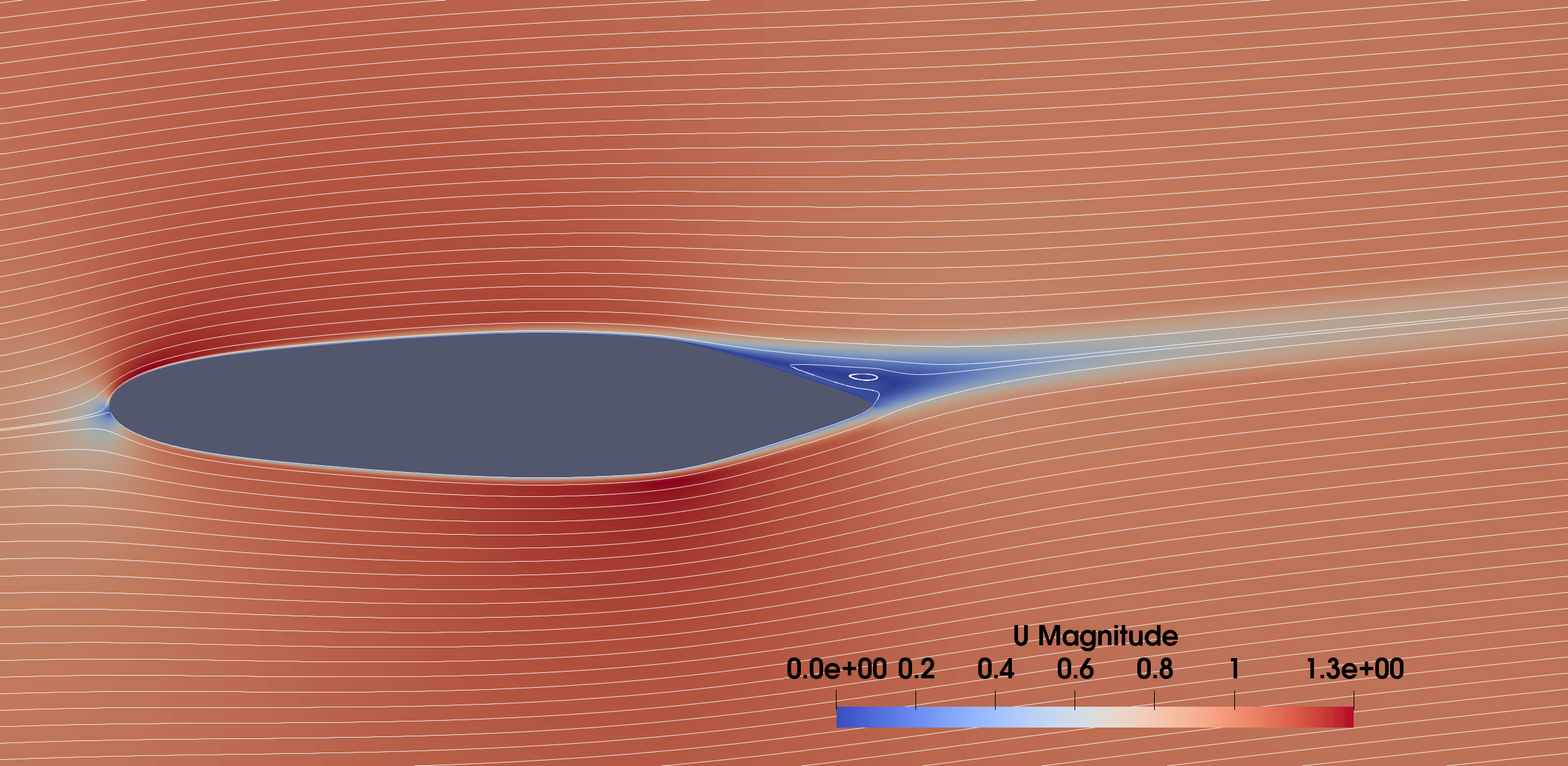}
    \caption{ \edit{Single, AoA = 5°}}
\end{subfigure}
\begin{subfigure}[t]{0.32\textwidth}
    \centering
    \includegraphics[width=\linewidth]{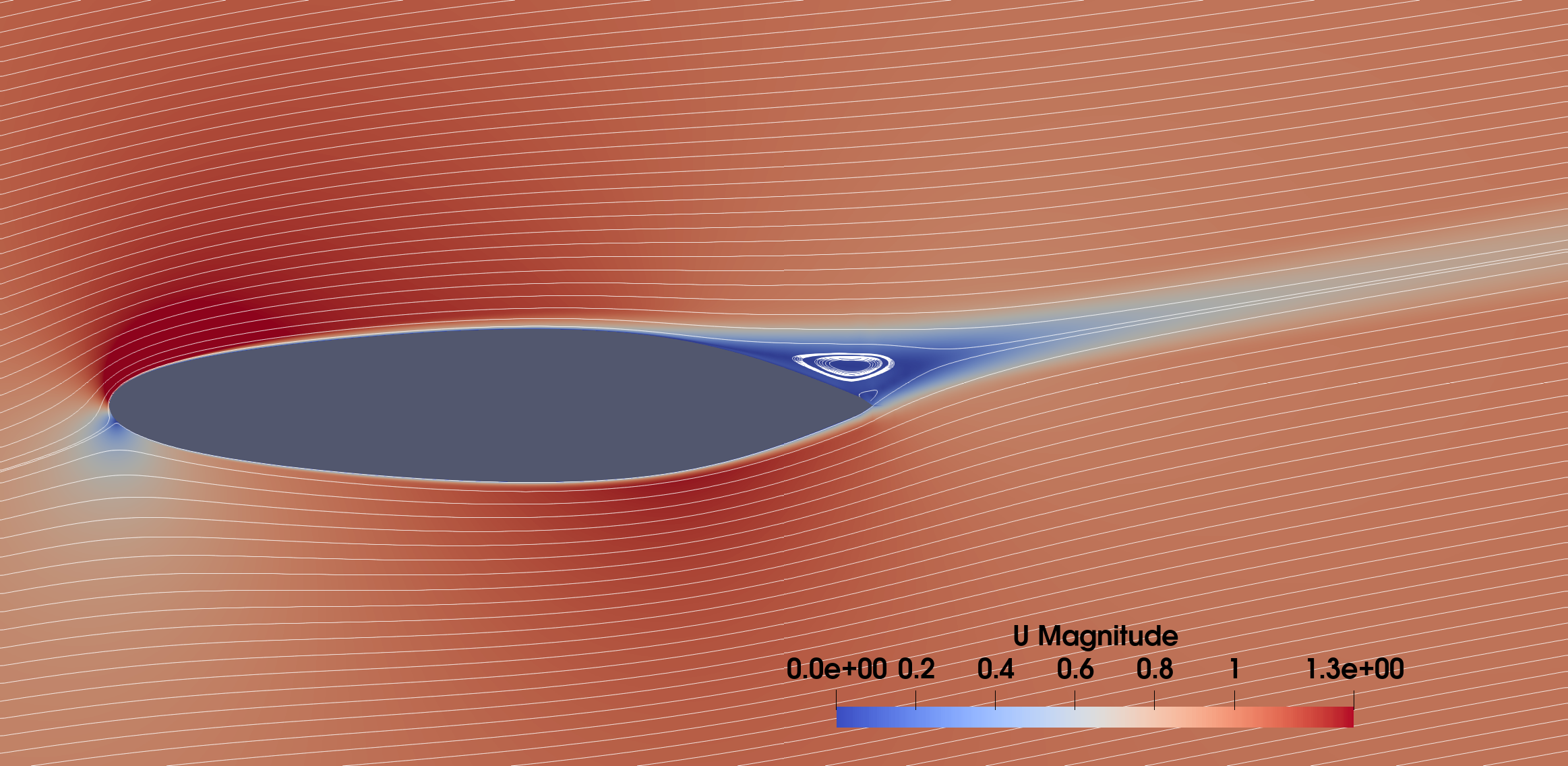}
    \caption{ \edit{Single, AoA = 10°}}
\end{subfigure}
\begin{subfigure}[t]{0.32\textwidth}
    \centering
    \includegraphics[width=\linewidth]{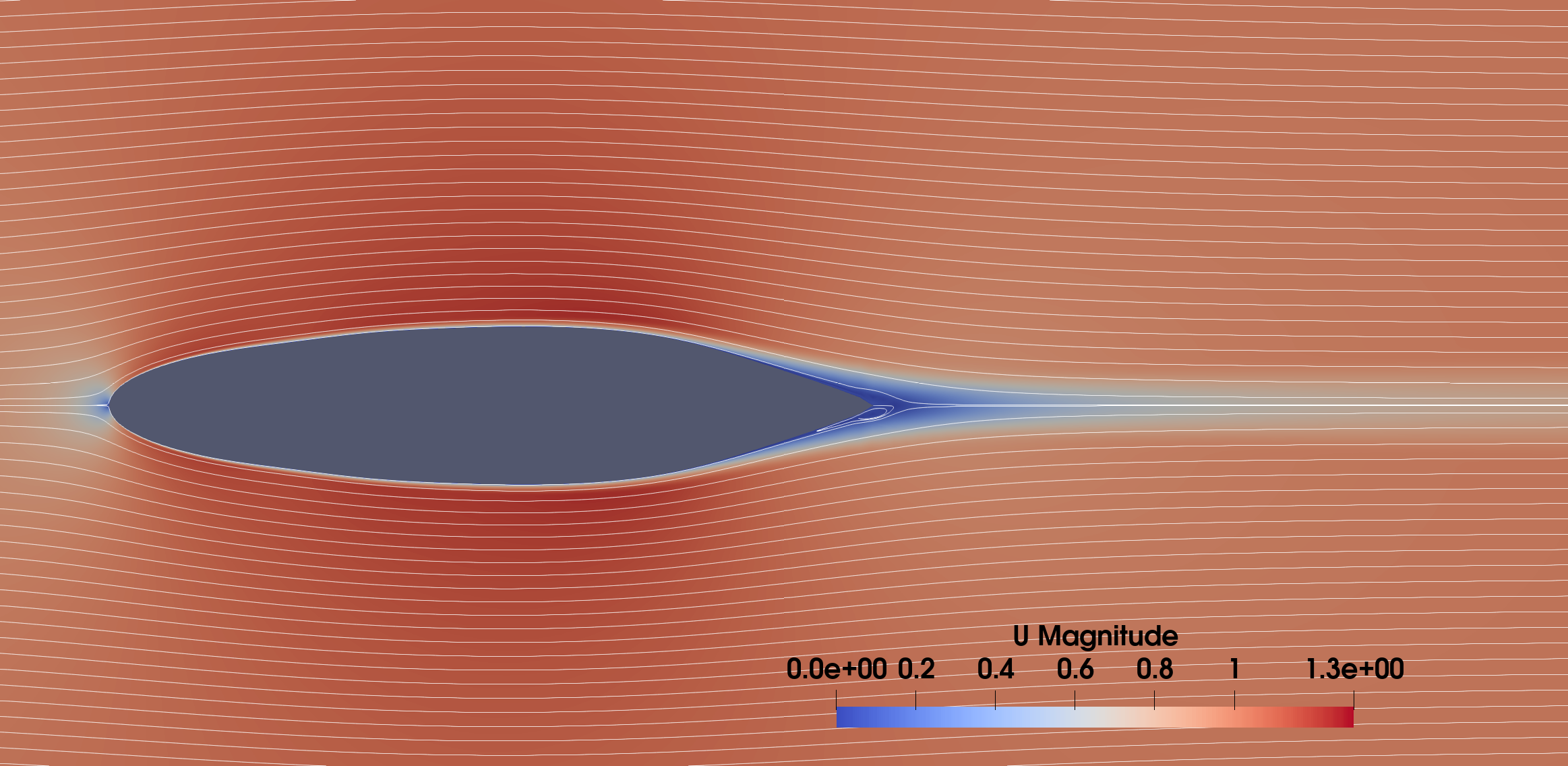}
    \caption{ Multiple, AoA = 0°}
\end{subfigure}
\begin{subfigure}[t]{0.32\textwidth}
    \centering
    \includegraphics[width=\linewidth]{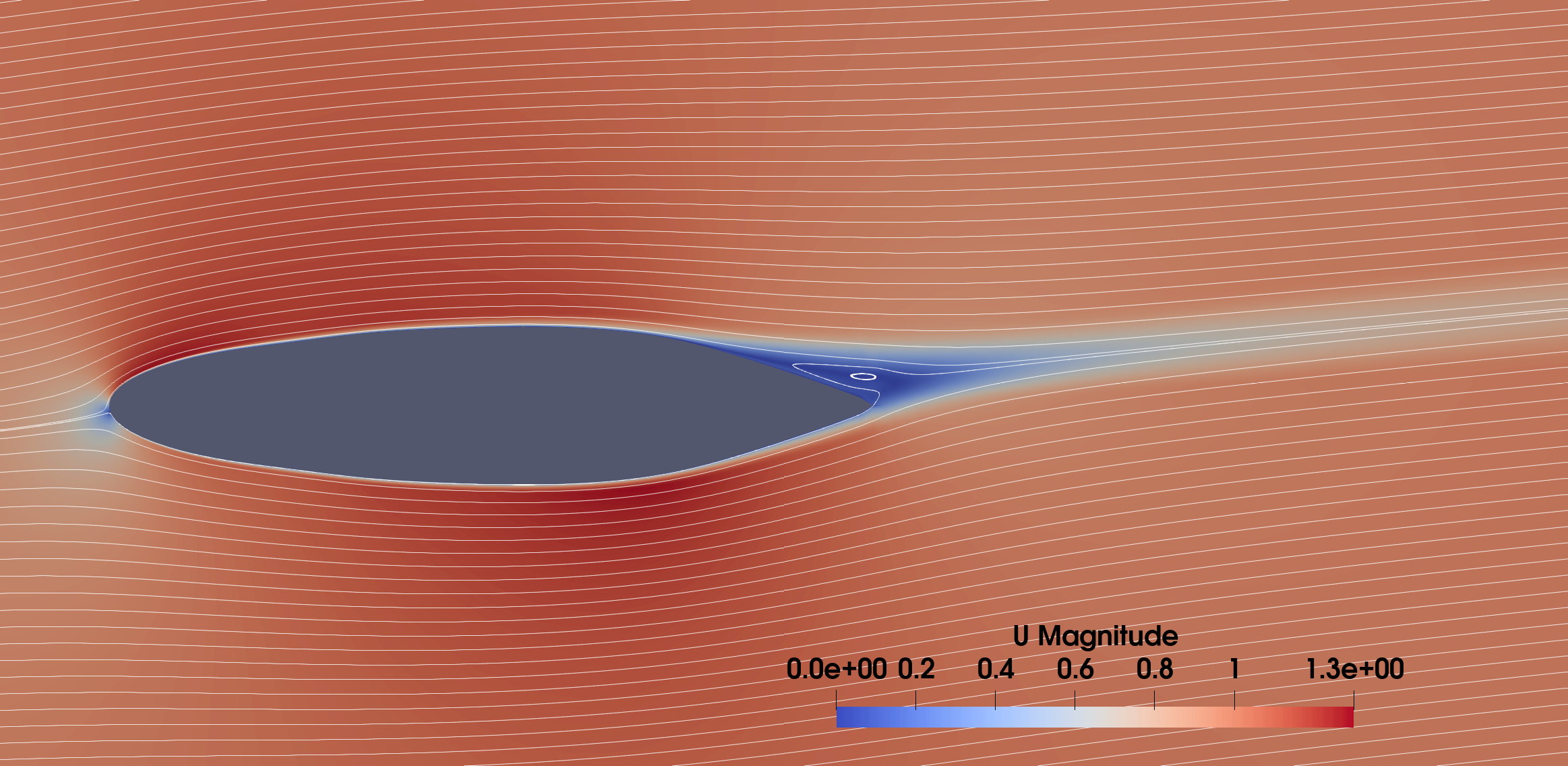}
    \caption{ Multiple, AoA = 5°}
\end{subfigure}
\begin{subfigure}[t]{0.32\textwidth}
    \centering
    \includegraphics[width=\linewidth]{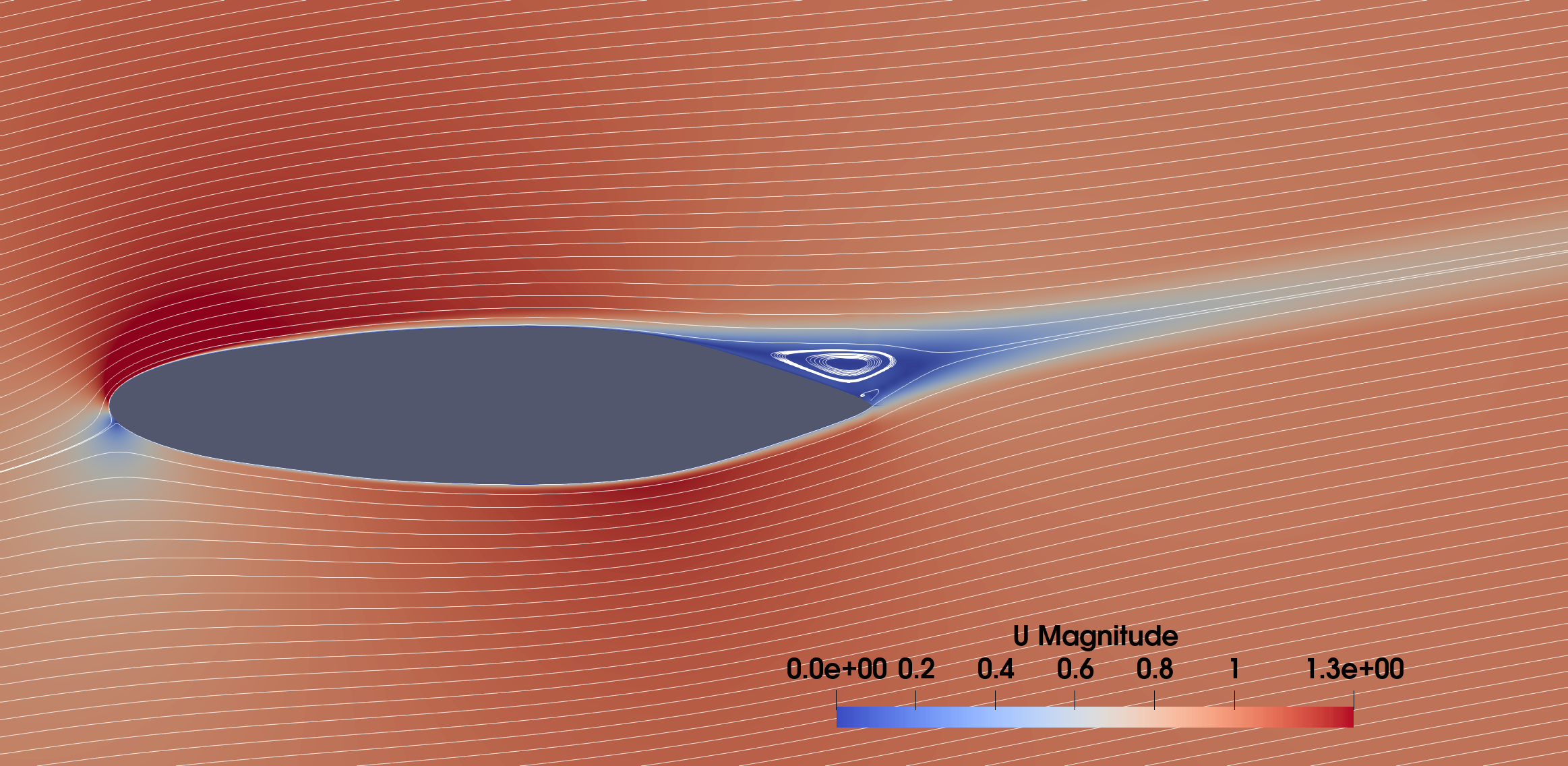}
    \caption{ Multiple, AoA = 10°}
\end{subfigure}
\caption{Velocity magnitude fields for airfoils at different angles of attack. “Single” denotes the airfoils identified using only one angle of attack ($0^\circ$, $5^\circ$, or $10^\circ$), while “Multiple” refers to the airfoil obtained through inverse shape optimization using three angles of attack. All simulations are performed with the Spalart–Allmaras (S–A) turbulence model, which is also employed in the adjoint-based shape optimization for airfoil identification.}
\label{fig:field_multiple}
\end{figure*}

\edit{We inversely obtained four airfoils using the wake signature from single AoA ($0^\circ$, $5^\circ$, or $10^\circ$) and three AoAs ($0^\circ$, $5^\circ$, and $10^\circ$).} The initial guess of the adjoint shape optimization is the classical NACA0012 airfoil. The FFD design variables of the NACA0012 airfoil are updated to minimize objective defined in \autoref{eq:obj} or \autoref{eq:obj_reply}. \autoref{fig:field_multiple} shows the flow fields of these airfoils, including the initial guess NACA0012, the target profile NACA16021, and inversely obtained airfoils based on single and multiple wake signatures. Note that the region where the wake signature is calculated is shown in \autoref{fig:domain} and extends longer downstream than the region shown here. Given the wake signature downstream and the drag and lift coefficients, the overall airfoil shape can be determined and the whole velocity field of the inversely determined airfoils is similar to that of the target airfoil. The airfoil obtained from the single wake signature deviates more from the target than that obtained from multiple wake signatures.

\begin{figure}[!h]
    \centering
    \includegraphics[width=0.8\linewidth]{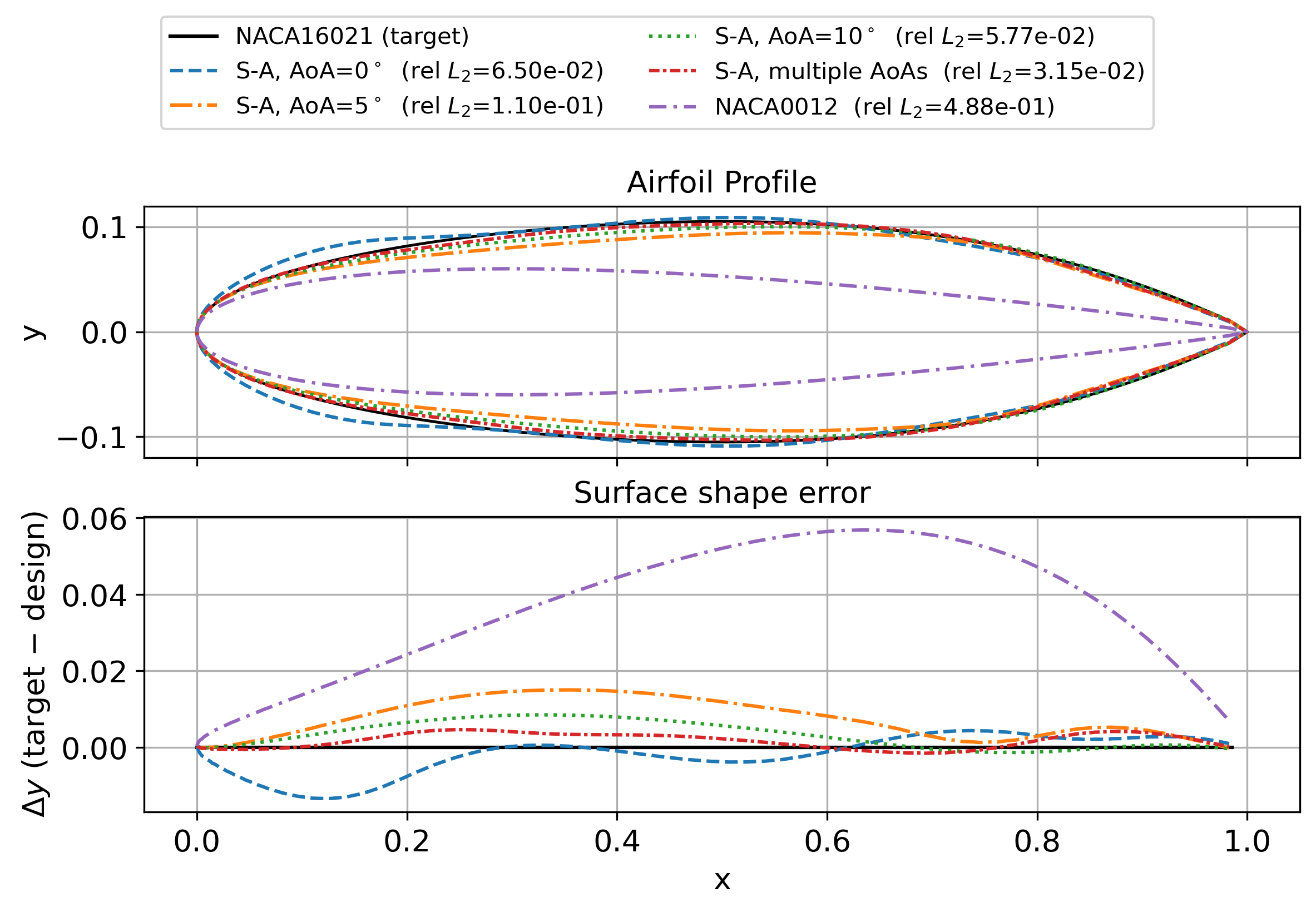}
    \caption{\edit{Comparison among the target, initial, and four inversely obtained airfoil profiles. The target profile is NACA16021, and the initial guess of optimization is NACA0012. The S-A model is used to calculate the wake signatures and determine the shapes. Four inversely obtained airfoils are based on either one or three angles of attack, respectively. The pointwise shape errors and relative $L_2$ errors of the profile are also shown.}}
    \label{fig:profile_multiple}
\end{figure}

\autoref{fig:profile_multiple} compares the airfoil profiles of the initial guess, the target, and \edit{four} inversely identified shapes. Although the airfoil identified from a single wake signature shows resemblance to the target NACA16021, it exhibits noticeably larger deviations than the one reconstructed using multiple wake signatures. \edit{Among the shapes identified from a single angle of attack, the $AoA=10^\circ$ case shows the best accuracy ($5.77\%$), and the $AoA=5^\circ$ case is the worst ($11.0\%$). The shape identified from three AoAs is the most accurate, whose relative $L_2$ error is $3.15\%$. All relative $L_2$ and $L_\infty$ errors of profiles are summarized in \autoref{tab:shape_errors}.}

\edit{\autoref{tab:error_forward} further summarizes the individual terms in the objective function for all airfoils, including the target NACA16021, the initial guess NACA0012, and all identified shapes. The reported quantities include the velocity-difference term and the drag and lift coefficients. Using the S-A turbulence model, the velocity-difference term $\Delta U^2$ is reduced by approximately three orders of magnitude in all cases, while both $C_D$ and $C_L$ are predicted much more accurately than for the initial NACA0012 shape. The ``Single S-A'' cases generally achieve higher accuracy at the angles of attack for which reference data are provided, but perform worse under unseen conditions, indicating an overfitting phenomenon. By contrast, the ``Multi S-A'' case yields lower velocity-field errors than the ``Single S-A'' cases because it incorporates information from all operating conditions during the optimization. This improved consistency across conditions also explains why the ``Multi S-A'' case achieves higher shape accuracy than the ``Single S-A'' cases.}
\begin{table}[h!]
  \centering
  \caption{\edit{Relative airfoil shape errors with respect to NACA\,16-021 target.}}
  \label{tab:shape_errors}
  \begin{tabular}{lcc}
    \toprule
    Case & rel.\,$L_2$ & rel.\,$L_\infty$ \\
    \midrule
    NACA0012 & 4.88e-01 & 5.42e-01 \\
    S-A, AoA=$0^\circ$ & 6.50e-02 & 1.28e-01 \\
    S-A, AoA=$5^\circ$ & 1.10e-01 & 1.43e-01 \\
    S-A, AoA=$10^\circ$ & 5.77e-02 & 8.10e-02 \\
    \textbf{S-A, multiple AoAs} & \textbf{3.15e-02} & \textbf{4.40e-02} \\
    \midrule
    S-A, AoA=$10^\circ$ & 5.77e-02 & 8.10e-02 \\
    \textbf{S-A, AoA=$10^\circ$, U only} & \textbf{2.47e-02} & \textbf{3.91e-02} \\
    S-A, AoA=$10^\circ$, Cd/Cl only & 1.14e-01 & 1.68e-01 \\
    S-A, AoA=$10^\circ$, smaller domain & 7.18e-02 & 9.36e-02 \\
    S-A, AoA=$10^\circ$, fewer FFD points & 6.38e-02 & 8.51e-02 \\
    \midrule
    \textbf{S-A, multiple AoAs} & \textbf{3.15e-02} & \textbf{4.40e-02} \\
    $k-\varepsilon$, multiple AoAs & 2.17e-01 & 2.33e-01 \\
    $k-\omega$ SST, multiple AoAs & 1.65e-01 & 2.41e-01 \\
    \bottomrule
  \end{tabular}
\end{table}

\begin{table}[h!]
  \centering
  \caption{\edit{Objective function values at convergence. $\Delta U^2$ is the velocity-field variance; $C_D$ and $C_L$ are the drag and lift coefficients of the optimized shape.}}
  \label{tab:error_forward}
    \scriptsize
\setlength{\tabcolsep}{3pt}  
  \begin{tabular}{l ccc ccc ccc}
    \toprule
    & \multicolumn{3}{c}{$\alpha = 0^\circ$} & \multicolumn{3}{c}{$\alpha = 5^\circ$} & \multicolumn{3}{c}{$\alpha = 10^\circ$} \\
    \cmidrule(lr){2-4} \cmidrule(lr){5-7} \cmidrule(lr){8-10}
    Case & $\Delta U^2$ & $C_D$ & $C_L$ & $\Delta U^2$ & $C_D$ & $C_L$ & $\Delta U^2$ & $C_D$ & $C_L$ \\
    \midrule
    NACA\,16021 & 0 & 2.61e-3 & 0 & 0 & 2.61e-3 & 4.54e-3 & 0 & 3.14e-3 & 2.82e-2 \\
    NACA\,0012 & 1.28e1 & 1.74e-3 & 3.44e-8 & 2.65e1 & 2.11e-3 & 4.51e-2 & 3.01e1 & 3.42e-3 & 8.62e-2 \\
    \midrule
    Single S-A $0^\circ$ & 2.20e-2 & \textbf{2.61e-3} & -5.31e-5 & 1.48e0& 2.67e-3& 5.47e-3& 2.33e-1& 3.29e-3& 3.34e-2\\
    Single S-A $5^\circ$ & 5.26e-2& 2.55e-3& -5.23e-5& 7.48e-2 & \textbf{2.66e-3} & 3.23e-3 & 5.29e-2& 3.18e-3& 2.86e-2\\
    Single S-A $10^\circ$ & 3.27e-1& 2.65e-2& -6.68e-5& 5.20e-1& 2.68e-3& 4.22e-4& 2.97e-2 & \textbf{3.15e-3} & \textbf{2.81e-2} \\
    Multi S-A & \textbf{2.18e-2} & 2.62e-3& -2.68e-5 & \textbf{4.32e-2} & 2.67e-3& \textbf{3.84e-3} & \textbf{2.78e-2} & \textbf{3.15e-3}& 2.88e-2 \\
    Multi $k$-$\omega$ SST & 3.08e-1 & 2.58e-3 & -1.02e-4 & 4.04e-1 & 2.83e-3 & 4.29e-3 & 5.57e-1 & 3.64e-3 & 3.69e-2 \\
    Multi $k$-$\varepsilon$ & 2.86e0 & 4.00e-3 & \textbf{-3.19e-6} & 1.23e1 & 4.34e-3 & 1.64e-2& 6.98e0 & 5.49e-3 & 3.78e-2 \\
    \bottomrule
  \end{tabular}
\end{table}

In summary, the inverse shape identification problem is inherently ill-posed, as evidenced by the noticeable discrepancies among the identified airfoil shapes. A practical way to mitigate this ill-posedness is to incorporate additional information.

\subsection{\edit{Effect of problem setup on identified shapes}}
\label{sec:effect}
\edit{
In this section, we will study the effect of the following factors on the accuracy of shape identification: (i) weights in the objective function \autoref{eq:obj}, (ii) the size of the domain on which the wake signature is computed, and (iii) the number of FFD control points in shape parameterization. All tests are based on shape identification using the single wake signature $AoA=10^\circ$, given the higher accuracy of $AoA=10^\circ$ over the other two cases using the single AoA wake signature, as shown in the first part of \autoref{tab:shape_errors}.
}

\begin{figure}[!h]
    \centering
    \includegraphics[width=0.8\linewidth]{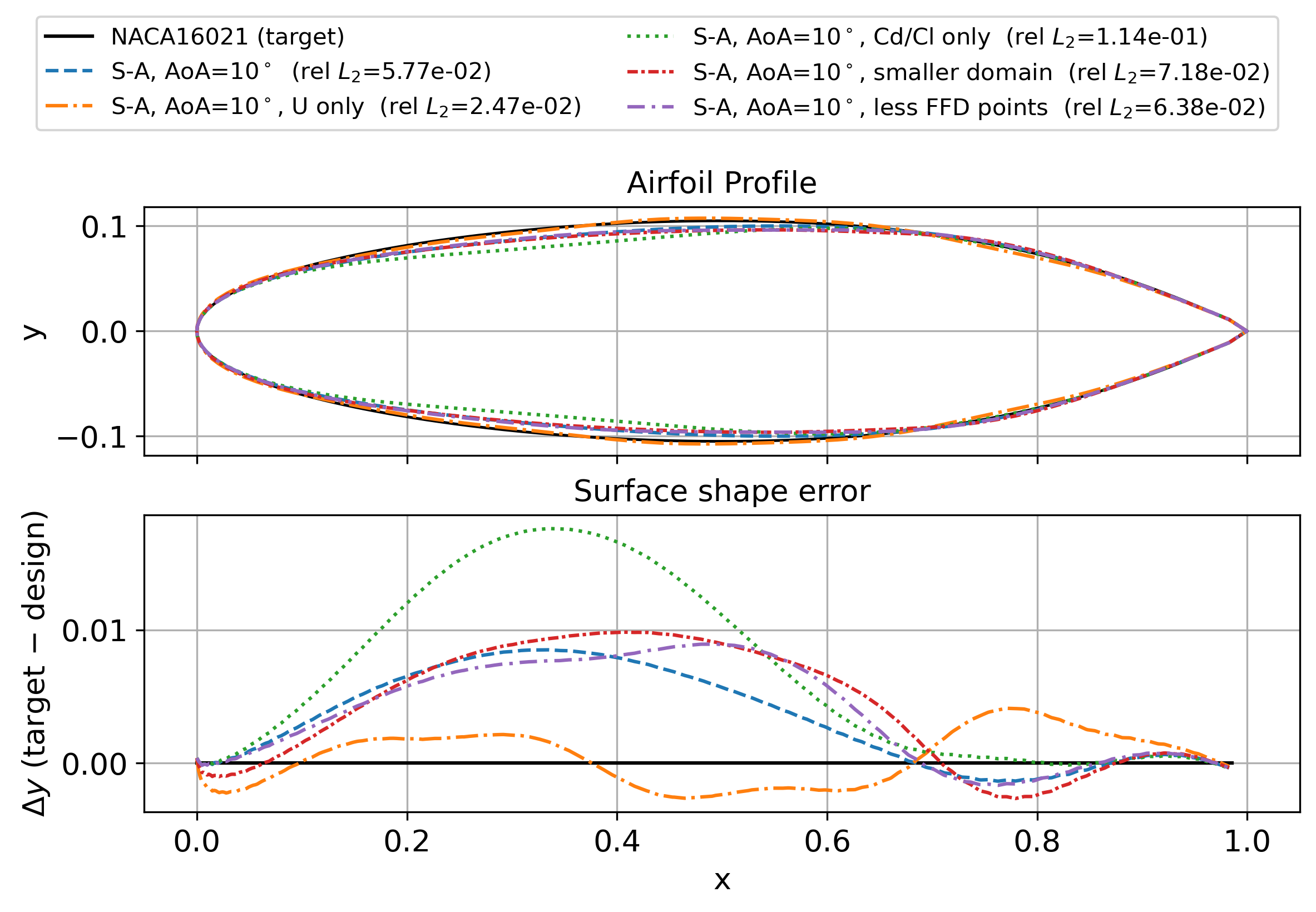}
    \caption{\edit{Comparison among the target, initial, and two inversely obtained airfoil profiles. The target profile is NACA16021, and the initial guess of optimization is NACA0012. The S-A model is used to calculate the wake signatures and determine the shapes. Two inversely obtained airfoils are based on one and three angles of attack, respectively.}}
    \label{fig:profile_effect}
\end{figure}

\edit{
\autoref{fig:profile_effect} compares the identified profiles using different setups. ``U only'' means setting $w_1=w_2=0$ in the objective (\autoref{eq:obj}) and only using the information from velocity field; ``$C_D/C_L$ only'' means setting $w_0=0$ and only using information from coefficients; ``smaller domain'' means instead of $\Omega (x\in[1.1,4.5], y\in[-0.5,0.5])$ as shown in \autoref{fig:domain}, a smaller domain $\Omega' (x\in[2.0,3.0], y\in[-0.5,0.5])$ is used to calculate the velocity difference in the objective function; and ``fewer FFD points'' means instead of have $2\times 10$ FFD control points as shown in \autoref{fig:FFD}, fewer $2\times 6$ points are used to parameterize the geometry. Relative $L_2$ and $L_\infty$ errors of all identified shapes are summarized in the second part of \autoref{tab:shape_errors}.
}

\edit{
Results show that changing the weights in the objective function affects the identified shapes largely. Using only the aerodynamic coefficients yields the largest error ($11.4\%$), whereas using only the velocity field yields even better results ($2.47\%$) than the baseline setup. One possible reason is that the optimization is overfitting the target coefficients that have non-zero discretization errors and have a stronger non-linear relationship with the geometric parameters. A smaller domain and fewer FFD points lead to a marginal difference in identified shapes, and even the pointwise shape errors follow a similar trend to the baseline setup. 
}

\subsection{Identifying shapes using consistent and inconsistent turbulence closures}\label{sec:closure}

After solving the inverse shape identification problem using multiple wake signatures, our focus shifts to examining the influence of turbulence-closure consistency within the optimization framework. Although the uncertainty associated with turbulence closures has been extensively investigated from the perspective of forward (functional) prediction, it has received comparatively little attention in the context of inverse problems.

\begin{figure}[h!]
    \centering
    \includegraphics[width=0.8\linewidth]{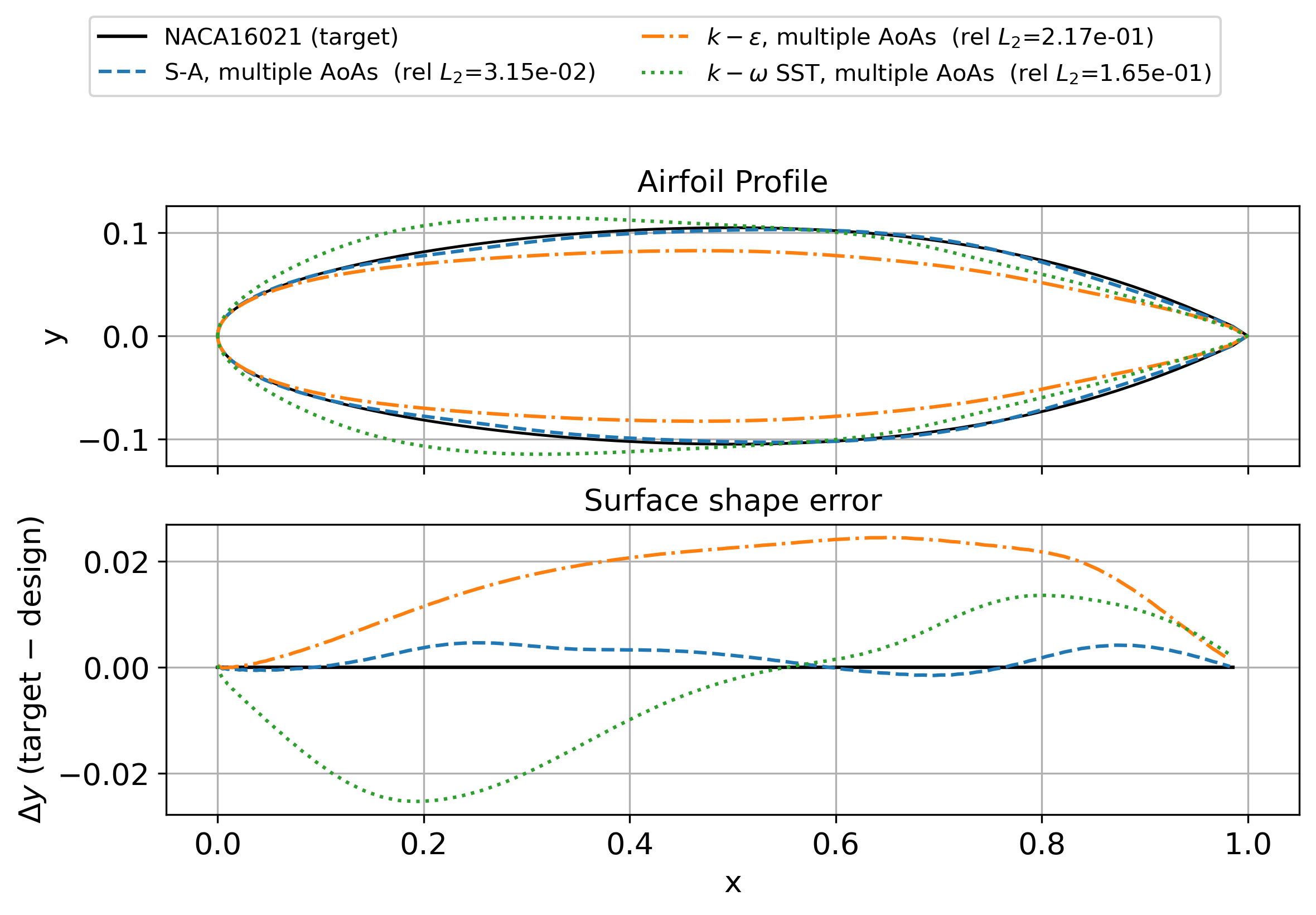}
    \caption{\edit{Comparison among the target and three inversely obtained airfoil profiles. The target profile is NACA16021, and its wake signature is calculated using the S-A model. Three inversely obtained airfoils are calculated using the S-A model, the $k-\varepsilon$ model, and the $k-\omega$ SST model, respectively, and using three angles of attack.}}
    \label{fig:profile_closure}
\end{figure}

\autoref{fig:profile_closure} compares the shapes of inversely obtained airfoils against the target NACA16021. The one obtained from the $k-\varepsilon$ model deviates the most \edit{($21.7\%$)} from the target, while the next is the one obtained from the $k-\omega$ SST model \edit{($16.5\%$)}. \autoref{tab:shape_errors} (the last part) quantitatively summarizes the shape deviation of the three inversely obtained airfoils using multiple wake signatures. The consistent closure (S-A model) shows the smallest deviation from the target airfoil, both in terms of $L_2$ and $L_\infty$ errors. \autoref{tab:error_forward} summarizes the converged objective terms of three turbulence closures. The consistent closure S-A model shows the lowest errors.

\subsection{\edit{Flow and sensitivity comparison among different closure models}}
\label{sec:sens}

\begin{figure}[h!]
    \centering
    \includegraphics[width=1\linewidth]{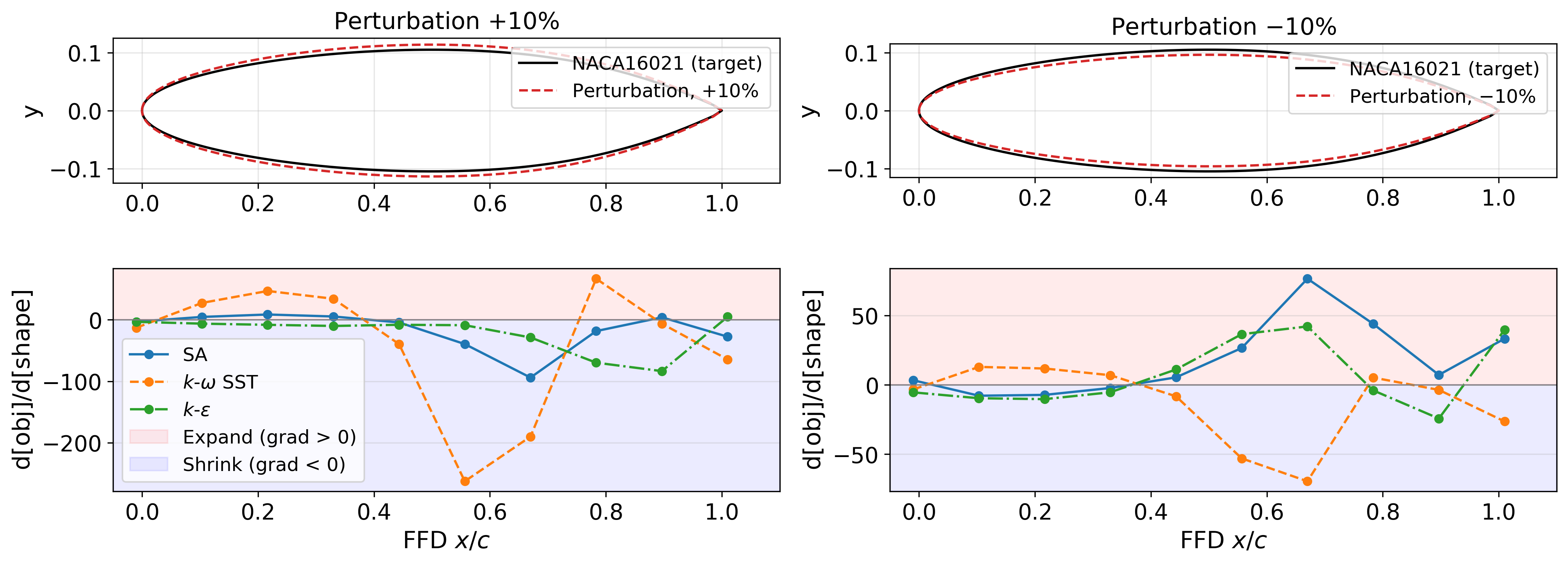}
    \caption{\edit{Perturbed NACA16021 airfoils and the corresponding shape sensitivities. The target NACA16021 airfoil is perturbed by symmetrically expanding or shrinking the FFD control points by $10\%$. For each perturbed shape, three turbulence models are used to evaluate the gradient of the objective function in \autoref{eq:obj} with respect to the vertical coordinates of the FFD control points. The resulting gradients indicate the direction and magnitude of the local shape modifications required to reduce the objective, i.e., to recover the unperturbed NACA16021 geometry.}}
    \label{fig:sens}
\end{figure}

\edit{
In this section, different turbulence models are compared in terms of both functional evaluation and shape sensitivity. The target airfoil, NACA16021, is perturbed by $\pm 10\%$, as shown in \autoref{fig:sens}. For each perturbed geometry, three turbulence models are used to evaluate the wake velocity mismatch $\Delta U^2$ with respect to the unperturbed NACA16021, together with the aerodynamic coefficients $C_D$ and $C_L$ at an angle of attack of $10^\circ$. }

\begin{table}[h!]
  \centering
  \caption{\edit{Wake mismatch and aerodynamic coefficients for perturbed NACA\,16-021 geometries at $\alpha=10^\circ$.}}
  \label{tab:sensitivity_obj}
  \begin{tabular}{llccc}
    \toprule
    Perturbation & Model & $\Delta U^2$ & $C_D$ & $C_L$ \\
    \midrule
    $+10\%$ & SA & 8.71e-01 & 3.25e-03 & 1.98e-02 \\
    $+10\%$ & $k$-$\omega$ SST & 6.25e+00 & 3.49e-03 & 7.37e-03 \\
    $+10\%$ & $k$-$\varepsilon$ & 7.24e+00 & 5.15e-03 & 2.88e-02 \\
    \midrule
    $-10\%$ & SA & 9.02e-01 & 3.06e-03 & 3.65e-02 \\
    $-10\%$ & $k$-$\omega$ SST & 7.31e-01 & 3.18e-03 & 2.44e-02 \\
    $-10\%$ & $k$-$\varepsilon$ & 5.83e+00 & 4.92e-03 & 3.93e-02 \\
    \bottomrule
  \end{tabular}
\end{table}

\edit{
\autoref{tab:sensitivity_obj} shows that the three closure models can produce substantially different functional evaluations for the same geometric perturbation. For the $+10\%$ perturbation, the S-A model predicts a wake mismatch of order unity, whereas the $k$-$\omega$ SST and $k$-$\varepsilon$ models predict values nearly one order of magnitude larger. Differences are also observed in the aerodynamic coefficients, with discrepancies in some cases reaching approximately $50\%$. We do not regard the S-A model as the ground truth and just show the difference among models. These variations already indicate a non-negligible model dependence at the level of the forward prediction.}

\edit{
A more important observation is obtained from the shape sensitivities shown in \autoref{fig:sens}. Positive gradients (red) indicate that local expansion would reduce the objective, whereas negative gradients (blue) indicate that local shrinkage is required. The gradients vary strongly along the chord, reflecting the nonlinear dependence of the flow response on the airfoil geometry. More importantly, they differ significantly across turbulence models, revealing a pronounced inconsistency in sensitivity prediction.}

\edit{
In this comparison, the S-A model is used as the reference closure, since the reference velocity field and force coefficients are also generated with S-A. Accordingly, its gradient distribution provides the self-consistent correction direction: for the positively perturbed geometry, the largest shrinkage is required near $x/c \approx 0.7$, while the opposite trend is observed for the negatively perturbed case. This location is close to the maximum-thickness region and therefore also corresponds to the largest local geometric perturbation. By contrast, the $k$-$\omega$ SST and $k$-$\varepsilon$ models predict markedly different gradient distributions. The relative $L_2$ difference with respect to the S-A gradient ranges from approximately $70\%$ to $250\%$. In several regions, the trends are not even reversed between the $\pm 10\%$ perturbations, suggesting correction directions that would lead to substantially different recovered shapes, consistent with the behavior shown in \autoref{fig:profile_closure}.}

\edit{
Overall, these results show that even when different closure models provide comparable forward predictions, they may still yield substantially different shape sensitivities. Such sensitivity inconsistency can directly lead to different inverse solutions and therefore should be considered explicitly in both inverse design and closure-model development. Since reliable gradient information is generally unavailable from experiments and is also difficult to obtain from high-fidelity simulations, establishing meaningful sensitivity-based benchmarks remains an important topic for future work.}

\section{Discussion}
The results presented above reveal that turbulence closures, though calibrated for forward predictive accuracy, may exhibit large inconsistencies in the inverse or adjoint context. This finding emphasizes that the traditional notion of “model accuracy” is incomplete: a closure model must also preserve the \emph{correct sensitivities} that govern design and control outcomes. The discrepancy observed here suggests that turbulence models can be “predictively correct but inversely inconsistent,” a property that may silently bias optimization-based design workflows across aerodynamic applications.

One implication of this work is the need for systematic metrics to quantify the \emph{sensitivity consistency} of turbulence models.  
Such metrics could be defined by comparing adjoint sensitivities or gradient fields across closures for a common flow configuration, e.g.,
\[
\mathcal{E}_{\text{sens}} = \|\nabla_\theta J_{M_1} - \nabla_\theta J_{M_2}\|_{L_2},
\]
which measures the discrepancy in design gradients between two closure models.  
Alternatively, surrogate models could be trained to predict how turbulence‐model discrepancies propagate into geometric uncertainty in inverse design.  
These approaches would transform closure inconsistency from a qualitative observation into a quantifiable property.

Recent data-driven or hybrid turbulence models primarily focus on improving mean‐flow predictions.  
The present findings suggest that future models should also be trained and validated for their adjoint or sensitivity behavior.  
Including gradient information in the training or validation process—e.g., by penalizing discrepancies in $\partial J / \partial \theta$—could lead to \emph{sensitivity-consistent} data-driven closures better suited for optimization and control tasks.

Beyond turbulence modeling, this study highlights the need for uncertainty quantification frameworks that explicitly incorporate model-form errors into inverse problems.  
In practical terms, combining multi-fidelity modeling, Bayesian calibration, or ensemble-based methods with adjoint solvers may help estimate the range of geometries consistent with both data and model uncertainty.  
In this sense, the proposed investigation serves as a first step toward a unified framework for \emph{inverse design under model discrepancy}.

\section{Summary}

Inverse design and optimization of aerodynamic shapes represent a fundamental and high–impact application of computational fluid dynamics (CFD) and turbulence closure modeling. In this work, we examined two critical aspects of this process: the \emph{ill-posedness} of the inverse shape identification problem and the \emph{consistency} of turbulence closures used within it. We first formulate an inverse problem where the airfoil shape is determined based on the information of the wake signature and aerodynamic coefficients. Then, we compared the shapes obtained from single and multiple AoAs, showing that the inverse problem is ill-posed and more information leads to lower errors in both forward function estimation and inverse shape identification, thus alleviating the ill-posedness of the inverse problem. We further demonstrated that turbulence-closure consistency strongly affects the outcome of inverse shape identification. The inconsistent turbulence closure can lead to errors more than one order of magnitude higher than those of the consistent closure. These discrepancies arise from differences in both forward flow and sensitivity (adjoint) fields, highlighting that a turbulence model must ensure not only accurate mean-flow predictions but also consistent gradients with respect to geometric variations. 

In summary, this study reveals that turbulence-closure inconsistency can fundamentally bias inverse design outcomes. Our findings underscore the need to evaluate closure models based on both their \emph{predictive fidelity} and their \emph{sensitivity consistency}. Future work should \edit{build realistic benchmarks for both flow fields and their sensitivities}, develop quantitative metrics to measure closure-induced sensitivity discrepancies, and explore data-driven or hybrid modeling strategies that incorporate adjoint information during training. Such developments will enable turbulence models that are not only predictive but also reliable in optimization and inverse-design applications.

\section{Acknowledgments}
This research was supported by the Defense Advanced Research Projects Agency (DARPA) under the Automated Prediction Aided by Quantized Simulators (APAQuS) program, Grant No. HR00112490526. 

\bibliographystyle{elsarticle-num-names}   
\bibliography{ref}
\end{document}